\documentclass[sigplan,screen]{acmart}

\copyrightyear{2026}
\acmYear{2026}
\setcopyright{cc}
\setcctype{by}
\acmConference[ASPLOS '26]{Proceedings of the 31st ACM International Conference on Architectural Support for Programming Languages and Operating Systems, Volume 1}{March 22--26, 2026}{Pittsburgh, PA, USA}
\acmBooktitle{Proceedings of the 31st ACM International Conference on Architectural Support for Programming Languages and Operating Systems, Volume 1 (ASPLOS '26), March 22--26, 2026, Pittsburgh, PA, USA}
\acmPrice{}
\acmDOI{10.1145/3760250.3762225}
\acmISBN{979-8-4007-2165-6/26/03}


\usepackage[]{hyperref}

\usepackage[utf8]{inputenc}
\usepackage[T1]{fontenc}
\usepackage{microtype}

\usepackage{xargs}
\usepackage{lipsum}
\usepackage{xparse}
\usepackage{xifthen, xstring}
\usepackage{xspace}
\usepackage{marginnote}
\usepackage{etoolbox}
\def\paperversion{normal}

\makeatletter
\patchcmd{\@addmarginpar}{\ifodd\c@page}{\ifodd\c@page\@tempcnta\m@ne}{}{}
\makeatother
\ifx\paperversion\paperversiondraft
  \makeatletter
  \if@ACM@journal
    \paperwidth=\dimexpr \paperwidth + 3.5cm\relax
    \oddsidemargin=\dimexpr\oddsidemargin + 0cm\relax
    \evensidemargin=\dimexpr\evensidemargin + 0cm\relax
    \marginparwidth=\dimexpr \marginparwidth + 3cm\relax
    \makeatletter
    \long\def\@mn@@@marginnote[#1]#2[#3]{%
      \begingroup
        \ifmmode\mn@strut\let\@tempa\mn@vadjust\else
          \if@inlabel\leavevmode\fi
          \ifhmode\mn@strut\let\@tempa\mn@vadjust\else\let\@tempa\mn@vlap\fi
        \fi
        \@tempa{%
          \vbox to\z@{%
            \vss
            \@mn@margintest
            \if@reversemargin\if@tempswa
                \@tempswafalse
              \else
                \@tempswatrue
            \fi\fi
              \rlap{%
                \if@mn@verbose
                  \PackageInfo{marginnote}{xpos seems to be \@mn@currxpos}%
                \fi
                \begingroup
                  \ifx\@mn@currxpos\relax\else\ifx\@mn@currxpos\@empty\else
                      \kern-\dimexpr\@mn@currxpos\relax
                  \fi\fi
                  \ifx\@mn@currpage\relax
                    \let\@mn@currpage\@ne
                  \fi
                  \if@twoside\ifodd\@mn@currpage\relax
                      \kern\oddsidemargin
                    \else
                      \kern\evensidemargin
                    \fi
                  \else
                    \kern\oddsidemargin
                  \fi
                  \kern 1in
                \endgroup
                \kern\marginnotetextwidth\kern\marginparsep
                \vbox to\z@{\kern\marginnotevadjust\kern #3
                  \vbox to\z@{%
                    \hsize\marginparwidth
                    \linewidth\hsize
                    \kern-\parskip
                    \marginfont\raggedrightmarginnote\strut\hspace{\z@}%
                    \ignorespaces#2\endgraf
                    \vss}%
                  \vss}%
              }%
          }%
        }%
      \endgroup
    }
    \makeatother
  \else
    \paperwidth=\dimexpr \paperwidth + 6cm\relax
    \oddsidemargin=\dimexpr\oddsidemargin + 3cm\relax
    \evensidemargin=\dimexpr\evensidemargin + 3cm\relax
    \marginparwidth=\dimexpr \marginparwidth + 3cm\relax
  \fi
  \makeatother
\fi

\usepackage[textsize=tiny]{todonotes}
\usepackage[normalem]{ulem}

\makeatletter
\font\uwavefont=lasyb10 scaled 652
\newcommand\colorwave[1][blue]{\bgroup\markoverwith{\lower3\p@\hbox{\uwavefont\textcolor{#1}{\char58}}}\ULon}

\newcommand\highlight[2]{{\color{#1}{\colorwave[#1]{#2}}}}
\makeatother

\makeatletter
\newcommand\InFloat[2]{\ifnum\@floatpenalty<0\relax#1\else#2\fi}
\makeatother

\newboolean{inComment}
\setboolean{inComment}{false}

\ifx\paperversion\paperversiondraft
\newcommand\createtodoauthor[2]{
  \def\tmpdefault{emptystring}
  \expandafter\newcommand\csname #1\endcsname[2][\tmpdefault]{
    \ifthenelse{\boolean{inComment}}{
      \PackageError{paper-template}{Comments in comments not supported}{}
    }{}\setboolean{inComment}{true}
    \def\tmp{##1}
    \InFloat{
        \smash{
	  \marginnote{
	    \todo[inline,linecolor=#2,backgroundcolor=#2,bordercolor=#2]
	      {\textbf{#1 (Figure):} ##2}
          }
        }
    }{\ifthenelse{\equal{\tmp}{\tmpdefault}} 
      {\todo[linecolor=#2,backgroundcolor=#2,bordercolor=#2]{\textbf{#1:} ##2}\ignorespaces}
      {\ifthenelse{\equal{##2}{}} 
        {\highlight{#2}{##1}}
        {\highlight{#2}{##1}\todo[linecolor=#2,backgroundcolor=#2,bordercolor=#2]
	  {\textbf{#1:} ##2}
	}
      }
    }
    \setboolean{inComment}{false}
  }
}
\else
\newcommand\createtodoauthor[2]{%
\expandafter\newcommand\csname #1\endcsname[2][]{##1}%
}%
\fi

\usepackage{minted}
\usepackage{etoolbox}

\makeatletter
\ifcsdef{minted@optlistcl@quote}
{
\ifwindows
  \renewcommand{\minted@optlistcl@quote}[2]{%
    \ifstrempty{#2}{\detokenize{#1}}{\detokenize{#1="#2"}}}
\else
  \renewcommand{\minted@optlistcl@quote}[2]{%
    \ifstrempty{#2}{\detokenize{#1}}{\detokenize{#1='#2'}}}
\fi

\newcommand{\minted@def@optcl@novalue}[2]{%
  \define@booleankey{minted@opt@g}{#1}%
    {\minted@addto@optlistcl{\minted@optlistcl@g}{#2}{}%
     \@namedef{minted@opt@g:#1}{true}}
    {\@namedef{minted@opt@g:#1}{false}}
  \define@booleankey{minted@opt@g@i}{#1}%
    {\minted@addto@optlistcl{\minted@optlistcl@g@i}{#2}{}%
     \@namedef{minted@opt@g@i:#1}{true}}
    {\@namedef{minted@opt@g@i:#1}{false}}
  \define@booleankey{minted@opt@lang}{#1}%
    {\minted@addto@optlistcl@lang{minted@optlistcl@lang\minted@lang}{#2}{}%
     \@namedef{minted@opt@lang\minted@lang:#1}{true}}
    {\@namedef{minted@opt@lang\minted@lang:#1}{false}}
  \define@booleankey{minted@opt@lang@i}{#1}%
    {\minted@addto@optlistcl@lang{minted@optlistcl@lang\minted@lang @i}{#2}{}%
     \@namedef{minted@opt@lang\minted@lang @i:#1}{true}}
    {\@namedef{minted@opt@lang\minted@lang @i:#1}{false}}
  \define@booleankey{minted@opt@cmd}{#1}%
      {\minted@addto@optlistcl{\minted@optlistcl@cmd}{#2}{}%
        \@namedef{minted@opt@cmd:#1}{true}}
      {\@namedef{minted@opt@cmd:#1}{false}}
}
\minted@def@optcl@novalue{customlexer}{-x}
}
{
}
\makeatother

\usepackage{tikz}
\usetikzlibrary{arrows}
\usetikzlibrary{shapes}

\tikzset{
  circledstyle/.style={
    shape=circle,
    #1,
    font=\tt\small,
    inner sep=0pt,
    outer sep=0pt,
    minimum size=1.2em,
    text=black
  }
}

\usepackage{multicol}
\usepackage{xcolor}
\usepackage{tabularx}
\usepackage{subcaption}
\usepackage{minted}
\usepackage{tikz}
\usetikzlibrary{tikzmark,decorations.pathreplacing,calc,decorations.markings}

\usepackage{enumitem}
\setlist[itemize]{noitemsep, nolistsep}
\setlist[enumerate]{noitemsep, nolistsep}
\setlist[description]{noitemsep, nolistsep}

\definecolor{colori}{RGB}{0,0,0}

\NewDocumentCommand\MyArrow{O{0pt}mmmO{out=150,in=210}}
{\begin{tikzpicture}[overlay, remember picture]%
  \draw [->,thick,line width=0.5pt,#4]%
    ( $ ({pic cs:#3}|-{pic cs:#2}) + (-#1,0.5ex) $ ) to[#5]%
    ( $ (pic cs:#3) + (-#1,0.5ex) $ );%
\end{tikzpicture}}

\usepackage{algorithm}
\usepackage{algpseudocode}

\usepackage{pifont}
\usepackage[nolist,nohyperlinks]{acronym}
\acrodef{IR}[IR]{\emph{Intermediate Representation}}

\usepackage{stackengine}

\definecolor{pairedNegOneLightGray}{HTML}{cacaca}
\definecolor{pairedNegTwoDarkGray}{HTML}{827b7b}
\definecolor{pairedOneLightBlue}{HTML}{a6cee3}
\definecolor{pairedTwoDarkBlue}{HTML}{1f78b4}
\definecolor{pairedThreeLightGreen}{HTML}{b2df8a}
\definecolor{pairedFourDarkGreen}{HTML}{33a02c}
\definecolor{pairedFiveLightRed}{HTML}{fb9a99}
\definecolor{pairedSixDarkRed}{HTML}{e31a1c}

\createtodoauthor{tg}{pairedThreeLightGreen}

\newcommand{\IOC}{{I}_{OC}}

\begin{document}
\title[The Configuration Wall]{The Configuration Wall: Characterization and Elimination of Accelerator Configuration Overhead}

\author{Josse Van Delm}
\authornote{Contributed equally to the paper}
\orcid{0000-0002-9503-403X}
\affiliation{%
\institution{KU Leuven}
\city{Leuven}
\country{Belgium}}
\email{josse.vandelm@kuleuven.be}

\author{Anton Lydike}
\authornotemark[1]
\orcid{0009-0001-9389-8512}
\affiliation{%
\institution{The University of Edinburgh}
\city{Edinburgh}
\country{United Kingdom}}
\email{anton.lydike@ed.ac.uk}

\author{Joren Dumoulin}
\orcid{0009-0005-0692-1227}
\affiliation{%
\institution{KU Leuven}
\city{Leuven}
\country{Belgium}}
\email{joren.dumoulin@kuleuven.be}

\author{Jonas Crols}
\orcid{0009-0004-8821-3939}
\affiliation{%
\institution{KU Leuven}
\city{Leuven}
\country{Belgium}}
\email{jonas.crols@student.kuleuven.be}

\author{Xiaoling Yi}
\orcid{0009-0003-3001-3611}
\affiliation{%
\institution{KU Leuven}
\city{Leuven}
\country{Belgium}}
\email{xiaoling.yi@kuleuven.be}

\author{Ryan Antonio}
\orcid{0000-0002-0286-4609}
\affiliation{%
\institution{KU Leuven}
\city{Leuven}
\country{Belgium}}
\email{ryan.antonio@kuleuven.be}

\author{Jackson Woodruff}
\orcid{0000-0003-2650-9596}
\affiliation{%
\institution{The University of Edinburgh}
\city{Edinburgh}
\country{United Kingdom}}
\email{jackson.woodruff@ed.ac.uk}

\author{Tobias Grosser}
\orcid{0000-0003-3874-6003}
\affiliation{%
\institution{University of Cambridge}
\city{Cambridge}
\country{United Kingdom}}
\email{tobias.grosser@cst.cam.ac.uk}

\author{Marian Verhelst}
\orcid{0000-0003-3495-9263}
\affiliation{%
\institution{KU Leuven}
\city{Leuven}
\country{Belgium}}
\email{marian.verhelst@kuleuven.be}

\begin{abstract}

Contemporary compute platforms increasingly offload compute kernels from CPU to integrated hardware accelerators to reach maximum performance per Watt.
Unfortunately, the time the CPU spends on setup control and synchronization has increased with growing accelerator complexity.
For systems with complex accelerators, this means that performance can be configuration-bound.
Faster accelerators are more severely impacted by this overlooked
performance drop, which we call the configuration wall.
Prior work evidences this wall and proposes ad-hoc solutions to reduce configuration overhead.
However, these solutions are not universally applicable, nor do they offer comprehensive insights into the underlying causes of performance degradation.
In this work, we first introduce a widely-applicable variant of the well-known roofline model to quantify when system performance is configuration-bound.  
To move systems out of the performance-bound region, we subsequently propose a domain-specific compiler abstraction and associated optimization passes.
We implement the abstraction and passes in the MLIR compiler framework to run optimized binaries on open-source architectures to prove its effectiveness and generality.
Experiments demonstrate a geomean performance boost of 2x on the open-source OpenGeMM system, by eliminating redundant configuration cycles and by automatically hiding the remaining configuration cycles.
Our work provides key insights in how accelerator performance is affected by setup mechanisms, thereby facilitating automatic code generation for circumventing the configuration wall.
\end{abstract}

\maketitle 
\renewcommand{\shortauthors}{Josse Van Delm et al.}

\section{Introduction}

\begin{figure}
    \centering
    \includegraphics[width=\linewidth]{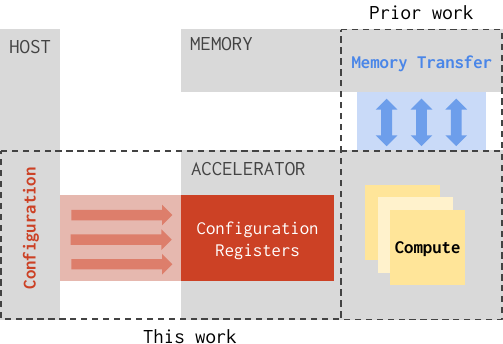}
    \caption{An accelerator's maximum attainable performance is not just limited by its memory bandwidth and compute resources, but also by its configuration interface.}
    \label{fig:architecture}
\end{figure}

\begin{figure}
    \centering
    \includegraphics[width=\linewidth]{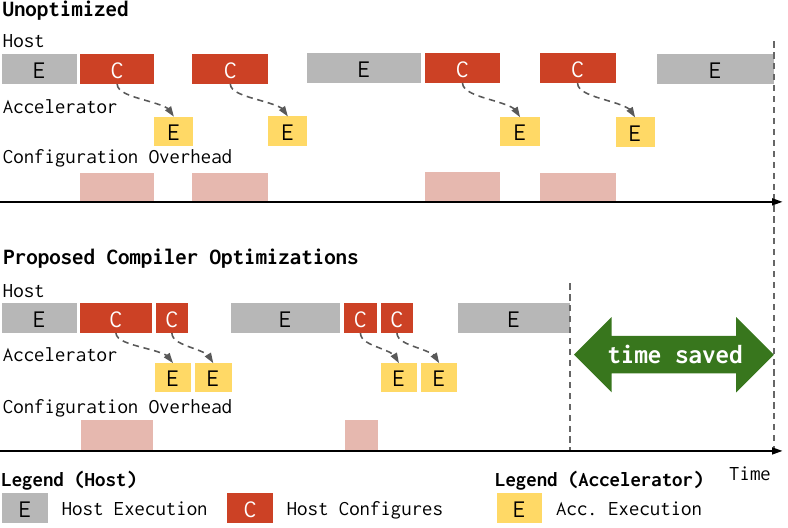}
    \tg{}
    \caption{This timeline of a typical program executed on a host CPU with dedicated hardware accelerator shows configuration overhead: cycles where CPU and accelerator are not performing any useful work.
    We present compiler techniques to eliminate this overhead, by making the configuration time shorter, and by overlapping it with other computations.}
    \label{fig:main_idea}
\end{figure}

With the end of Dennard scaling~\cite{baccarani2005generalized,dennard2007design} and traditional CPU
scaling trends~\cite{danowitz2012cpudb,horowitz2014computing}, computer architects are turning to
hardware accelerators to support emerging specialized workloads such as neural network inference.
Specialized fixed-function accelerators offer thousands of times
more performance than general-purpose processors~\cite{crypto_interface, crypto_fhe, crypto_facv,rpc_optimus_prime, rpc_zerializer, rpc_morpheus, rpc_protocolbuffer_accelerator, opengemm, gemmini, snitch, axi4mlir, redmule, nvidiah100, amx, nvdla, dl_riscv_multiprecision, dl_elastic_simd, dl_sssr, dl_hwacha, dl_piperench, dl_marsellus, dl_siracusa, dl_riscv_esp, neuromorphic_neurorvcore, robotics_racod} and
promise to overcome the challenges that computer architects
face in delivering next-generation high-performance compute.

However, hardware accelerators can hit the \textit{configuration wall}.
Time spent configuring the accelerator is not spent productively by either host
or accelerator. We call this the \emph{configuration overhead}.
Figure~\ref{fig:architecture} shows how accelerators typically
integrate into processor systems: they execute compute kernels on data directly coming from memory,
after being configured by a host processor.
The configuration wall characterizes an important trade-off:
Adding configuration options increases usefulness, but every added configuration option also directly reduces the achievable performance without proper optimizations.
As a result, a more reconfigurable accelerator may result in the system performing worse as a whole.
Existing theoretical models do not characterize this trade-off well and existing
compilers are unable to perform the needed optimizations.

This paper introduces a new roofline model to formalize the configuration
wall and enable compiler optimizations to optimize around it.
Our roofline models the \emph{``operation-to-configuration intensity''} against the maximum
\emph{``configuration bandwidth''}, and allows us to determine when accelerators hit the 
configuration wall. 
We use our roofline model to motivate two key compiler
optimizations: \emph{configuration deduplication} and \emph{configuration-computation
overlapping}.
We believe that both our model and proposed optimizations apply to a wide range of accelerators, 
regardless of domain, host processor, or interconnection with the rest of the compute system.

To achieve this, we introduce an MLIR dialect, \texttt{accfg},
which tracks configuration states and enables compilers to move
programs out of the configuration-bound region. Figure~\ref{fig:main_idea} shows
how we use \texttt{accfg} to optimize programs: by overlapping
configuration with program execution when the accelerator supports
it and by reducing the amount of configuration required
when the compiler can prove that certain parts of the configuration does not change.
We specifically showcase this in the paper on OpenGeMM~\cite{opengemm}, and Gemmini~\cite{gemmini}:
two matrix-multiplication accelerators which both are tightly coupled to a distinct RISC-V host CPUs 
which control the accelerators through custom instruction sequences.
All our work is open source and available on GitHub\footnote{
    \url{https://github.com/kuleuven-micas/snax-mlir}
}.
These compiler optimizations enable a geomean speed-up of
2x on concurrent configuration systems. 
Our contributions are:
\begin{enumerate}
    \item A new roofline model to assess the configuration wall, that
    places configuration overhead in relation to
    other well-known performance bottlenecks
    (\autoref{subsec:the_configuration_roofline})
    \item A compiler abstraction --- \texttt{accfg}
    --- to track configuration state and to effectively move systems out of the configuration bound region (\autoref{sec:implementation})
    \item An evaluation of our abstraction showing speed-up as predicted by our roofline model (\autoref{sec:evaluation})
\end{enumerate}

\section{Background}
\label{sec:probing_configuration_overhead}
To understand how the configuration wall limits performance, we outline additional information about accelerators and explain how they are configured.
To keep the discussion in this paper general, we mainly discuss systems that are tightly coupled to a CPU core which controls and synchronizes the accelerator through simple registers without the presence of an operating system (OS).
For such systems, we also discuss two configuration schemes: sequential and concurrent configuration.
More complex accelerators will typically perform the same operations albeit through more layers of hardware protocols and/or software abstraction layers.
Finally, we use Gemmini~\cite{gemmini} 
as a concrete example.

\subsection{Configuration in Accelerators}
\label{sec:configuration_introduction}

\MyArrow[1.5em]{start1}{end1}{colori}[out=180,in=180]
Typically, accelerators operate
in \textit{macro-operations}, i.e.,
a large algorithm step is executed as
a single instruction.  Many accelerators
rely on a configuration-based
strategy to enable a mix of flexibility
and performance:
\begin{enumerate}
    \item\tikzmark{end1}\textbf{Configure} accelerator - write configuration registers
    \item\textbf{Launch} computation - e.g., write to a launch register
    \item\tikzmark{start1}\textbf{Await} completion - e.g., poll a status register
\end{enumerate}
\smallskip

While each accelerator's semantics are different, the parameters sent to the accelerator during configuration share many similarities with regular CPU instructions.
Accelerators need to specify source and destination, typically expressed as memory addresses, the location of dedicated registers,
and/or other scalar parameters of the operation that are to be carried out.
However, as macro operations are much more complex than normal CPU instructions, the number of parameters is typically much larger, and the number of bits per parameter much higher than for processor instructions.

Because more configuration bytes have to be conveyed from the host to the accelerator, more cycles are required to fully configure an accelerator to perform the desired macro-operation.
This overhead has to be carefully managed, as it does not contribute to the actual calculation of the macro-operation that the accelerator was designed for.

As more cycles are typically required to fully configure an accelerator, getting the configuration mechanism right is crucial to achieving good accelerator performance.
As an example, assume a fictional accelerator that can process 100 elements per cycle. However, when considering that we need 3 cycles prior to every launch to configure it, we can see that the accelerator can only achieve an average throughput of 25 elements per cycle in actual workloads.

\subsection{Sequential vs. Concurrent Configuration}
\label{subsec:seqvsconc}

To minimize the effect of configuration overhead in accelerators, different configuration schemes can be used at the cost of extra complexity in both hard- and software.

The simplest scheme to configure an accelerator --- \textit{``sequential configuration''}, involves the host and accelerator taking turns to execute the program;
the host is stalled after issuing a launch of the accelerator.
Once the accelerator is finished, the host can configure the accelerator again or continue normal program execution~\cite{amx, gemmini}.

In a second scheme --- \textit{``concurrent configuration''}, the host is not stalled after issuing a launch of the accelerator, but it can already configure the next accelerator invocation, and/or continue processing while the accelerator is running~\cite{nvidiah100,redmule,nvdla}.
This gives the programmer the ability to overlap accelerator computation and configuration through \textit{synchronization barriers}.
This scheme additionally requires extra hardware, such as registers that stage new configurations while retaining old ones and other facilities to operate the barriers, making both hard- and software more complex. Other configuration mechanisms exist (see \autoref{sec:conclusion_outlook}), but are not discussed in depth in this work.

Understanding the supported scheme of each accelerator is crucial to determine how large the configuration overhead is, and what can be done about it in compiler optimizations.

\subsection{Data Movement $\neq$ Configuration Overhead}

Macro-operations can usually only be performed after all input data is in the right memory or register, but we do not consider such data movement part of the configuration.
Nonetheless, this data movement is crucial: if this transfer takes a long time, the accelerator's processing elements are sitting idle waiting for data.
As such, the accelerator will operate below peak performance, not because it is configuration-bound, but because it is memory-bound. 
In \autoref{subsec:3droofline}, we explain why this is orthogonal to accelerator configuration.

\subsection{Gemmini}
To make the previously discussed background concrete, we discuss Gemmini~\cite{gemmini} in detail.
Gemmini is a fully open-source processor architecture with a 64-bit Rocket~\cite{rocket} CPU and a systolic array to accelerate matrix multiplication.
The systolic array is operated by a controller that is configured through hardware registers that the rocket CPU writes to using custom instructions. 
These custom ``RoCC'' instructions can carry more information than a typical RISC-V instruction.
Gemini is sequentially configured: it cannot be reconfigured by the Rocket host CPU while it is processing.

Gemmini features two types of instruction sequences: fine-grained ones that spatially unroll the matrix multiplication on the systolic array, and coarse-grained instructions that also temporally unroll it.
As an example, we show some argument fields of the coarse-grained \texttt{gemmini\_loop\_ws} instruction sequence in \autoref{tab:gemmini_fields}.
These fields control various options for a weight-stationary matrix multiplication 
$\mathbf{C} = \mathbf{A}\mathbf{B} + \mathbf{D}$: the addresses and sizes of the individual matrices, selection of an activation function to be applied on the output, and whether the input should be accessed in a strided or transposed manner.
Since many of the fields are smaller than typical C word lengths, the CPU has to bit-pack these values, as shown in \autoref{fig:gemmini_setup}, an excerpt of Gemmini's C API.
Gemmini does not have dedicated launch instructions, instead the last instruction in the sequence implicitly launches it. We call this a \emph{``launch-semantic''} instruction.

The information presented here is crucial to derive the configuration overhead for Gemmini in \autoref{sec:gemmini_example_numbers}. 

\begin{table}[]
    \centering
    \caption{Some fields of the \texttt{gemmini\_loop\_ws sequence}, which performs a weight-stationary matrix multiplication $\mathbf{C} = \mathbf{A}\mathbf{B} + \mathbf{D}$, with their meaning and bitwidth}
    \resizebox{\columnwidth}{!}{%
    \begin{tabular}{l l l}
        \textbf{Field}                & \textbf{Meaning}      & \textbf{Bits} \\ \hline
                \texttt{A}, \texttt{B}, \texttt{D}, \texttt{C}       &  Address in main memory to matrices & 64 \\
                \texttt{I}, \texttt{J}, \texttt{K}        &  Sizes of the matrices & 16 \\
                \texttt{pad\_\{I,J,K\}}        &  Padding applied to sizes of the matrices & 16 \\
                \texttt{stride\_\{A,B,D,C\}}       &  Row strides to access matrices in memory & 64 \\
                \texttt{act} & Activation function application on output & 6  \\
                \texttt{\{A,B\}\_transpose} & Whether input matrix is transposed & 1 \\
        \hline
    \end{tabular}
    }
    \label{tab:gemmini_fields}
\end{table}

\section{Motivation}
In this section, we 
show why current compilers are unable to optimize code that interacts with external accelerators and demonstrate why purpose-built compiler abstractions are necessary to properly optimize configurations.

\subsection{Compilers do not Understand Configuration} \label{subsec:the_volatile_problem}

\begin{listing}
    \centering
    \includegraphics[width=\linewidth]{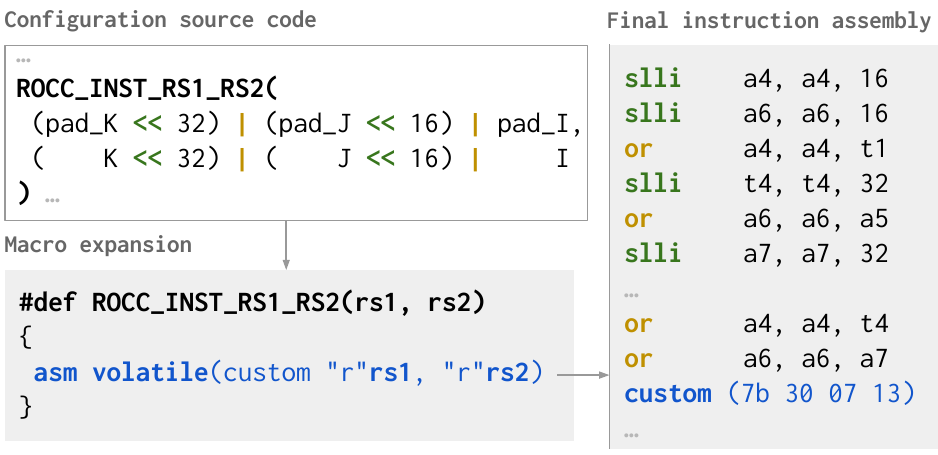}
    \caption{A cleaned up and truncated version of Gemmini's accelerator configuration and dispatch code will appear to the compiler as an opaque series of volatile assembly statements, providing little information in regard to their semantics, therefore preventing any and all compiler optimization.}
    \label{fig:gemmini_setup}
\end{listing}

Most programming languages were never specifically designed for accelerator configuration code - even C, a common systems language. This makes it difficult to optimize such configuration code for most compilers.

We identify three main reasons that prevent compilers from optimizing accelerator configuration code.
Firstly, accelerator configuration code is highly specific. The number and semantics of configuration parameters varies greatly among accelerators. Furthermore, many accelerators use custom instructions which typically require inline assembly. However, inline assembly renders emitted instruction sequences target-specific and mostly opaque to the compiler.
Secondly, configuration has to be performed in the right order. A compiler can not just reorder or omit configuration sequences, as launching an accelerator before it is configured may cause faults. This is especially important for accelerators which are launched implicitly after a certain parameter is set. Such launch-semantic configuration instructions always need to be executed last.
Thirdly, accelerator configuration is typically viewed as external register state. Compilers can track the state of general purpose registers, but not of special purpose ones such as accelerator configuration registers.
To inform the compiler that these registers are outside its control, developers have to use the \texttt{volatile} keyword. 

In practice these reasons typically force people to use volatile inline assembly, which fully prevents the compiler to optimize any accelerator configuration code.
Volatile inline assembly is indeed target-specific, always emitted, and emitted in the specified order (e.g., in Gemmini's setup sequence in \autoref{fig:gemmini_setup}).
This means that optimization to accelerator configuration code has to happen manually.

\subsection{The Promise of Compiler Optimizations}\label{sec:intro_optimisations}

Compiler's inability to automatically optimize accelerator configuration sequences does not prevent programmers from manually optimizing their accelerator invocations.
In fact, we can think of two key optimizations that can be implemented for most accelerators:

\textbf{Redundant Setup Elimination} which exploits the fact that configuration registers retain the configured values. If a configuration register already holds a given value from a previous setup, the second write to the register can be safely removed. This trick is most useful when the difference between repeated accelerator invocations is small, e.g., when doing tiled computations.

\textbf{Configuration-Computation Overlap} leverages extra hardware capabilities in concurrent configuration systems (see \autoref{subsec:seqvsconc}). These added capabilities allow the programmer to schedule code to run while the accelerator is running, which can hide most -- if not all -- of the setup overhead behind the accelerator's run-time. 

Performing these two optimizations manually is so problem- and accelerator-specific that they are impractical and therefore error-prone.
Both optimizations become increasingly difficult to perform with more complex programs, making a compiler-based approach highly desirable. However,
since compilers don't understand configuration, a more potent abstraction is needed.

In order to understand how these two optimizations impact configuration overhead, and to determine
how beneficial they are, we need an effective model that can explain and predict observed performance. For this, we introduce the configuration roofline.

\section{The Configuration Roofline Model} \label{subsec:the_configuration_roofline}

In this section, we introduce the novel \emph{Configuration Roofline Model}, which we use to estimate the impact of configuration overhead on overall performance.
We start off by reminding the reader about the existing processor roofline model\footnote{\label{procrooflinename}We specifically refer to this as the ``\emph{processor} roofline model'' to distinguish it from the adapted roofline model we introduce later in the paper.}, and then introduce our novel extension --- the configuration roofline model --- for both sequential and concurrently configured accelerators. We also explore how configuration-related calculations affect this model.
Finally, we conclude this section by looking into the configuration overhead of Gemmini and predicting the impact of the optimizations motivated in \autoref{sec:intro_optimisations}.

\subsection{Background: Processor Roofline Model}

A well-established model for visualizing and assessing performance bottlenecks is \textit{the processor roofline model}~\cite{rooflinemodel}\footref{procrooflinename}.
This model allows to see if a workload's performance on a specific processor (or accelerator) is memory or compute bound.
Assuming concurrent compute and memory transfers, it uses:

\textbf{Peak Processor Performance} $P_{\text{Peak}}$\footnote{Sometimes this is referred to as Peak Floating-point Performance \cite{rooflinemodel}, but some processors (or accelerators) don't support floating point operations, so we use the more general term Peak Processor Performance here.} is defined as the number of operations the processor's (or accelerator's) datapath supports per unit of time (cycles or seconds) e.g., in floating point operations per second (FLOPs).

\textbf{Operational Intensity $I_{\text{Operational}}$} of an algorithm refers to the number of operations that have to be executed per byte of data access in ops/byte.

\begin{figure}
    \centering
    \includegraphics[width=\linewidth]{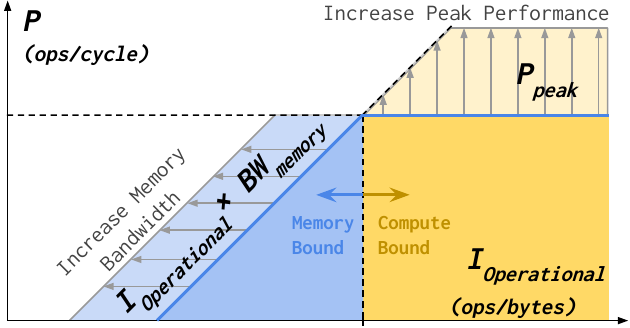}
    \caption{The processor roofline model, proposed in \cite{rooflinemodel}, can be used to quickly check whether an algorithm with operational intensity ${I}_\text{Operational}$ is compute-bound or memory-bound on a multi-core processor with peak performance $P_{\text{Peak}}$ and peak memory bandwidth ${BW}_{\text{Memory}}$.}
    \label{fig:regular_roofline_model}
\end{figure}

\textbf{Peak Memory Bandwidth ${BW}_{\text{Memory}}$} is defined as the maximum bandwidth that can practically be sustained for the memory~\cite{rooflinemodel}.
Now the maximum attainable performance $P_{\text{Attainable}}$ on a multi-core processor (or parallel accelerator) with peak performance $P_{\text{Peak}}$ and peak memory bandwidth ${BW}_{\text{Memory}}$ for a given algorithm with a certain operational intensity ${I}_\text{Operational}$ is plotted in \autoref{fig:regular_roofline_model} and given by:

\begin{equation}
P_{\text{Attainable}} = \min{
\begin{cases}
P_{\text{Peak}} \\
{BW}_{\text{Memory}} \times I_{\text{Operational}}
\end{cases}}
\label{eq:regular_roofline}
\end{equation}

Even though this model was mainly presented to perform performance analysis on multicore CPU architectures, it has been successfully extended and applied in many different systems.
Examples include extensions for cache \cite{cache_roofline}, distributed systems \cite{distributed_roofline}, GPUs \cite{gpumicrobenchmark, nvidiagpu, amdgpu} and TPUs \cite{tpu}.

\subsection{Concurrent Accelerator Configuration Roofline}

\begin{figure*}
    \centering
    \includegraphics[width=\linewidth]{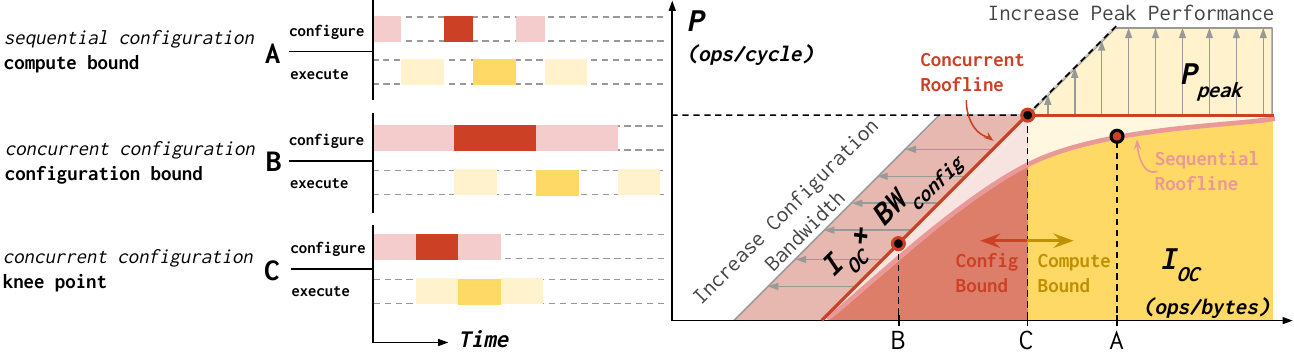}
    \caption{We propose the configuration roofline plot, which shows the maximum performance for systems with certain configuration capabilities.
    Sequential configuration systems and concurrent configuration systems are bound in performance by the light red and dark red lines respectively.
    The knee point in the middle divides the plot in two distinct regions, the \textit{configuration bound} on the left and \textit{compute bound region} on the right.
    On the left, timeline visualizations for either configuration capability are given in the configuration bound region (B), knee point (C) and compute bound region (A).}
    \label{fig:adapted_roofline_model}
\end{figure*}

Mirroring the classical roofline model, we propose a complementary \textit{configuration roofline} model. 
We mainly discuss this model for the host-controlled accelerators discussed in \autoref{sec:probing_configuration_overhead}, but just like the classical roofline model we believe this model can be applied to other systems.
For this model we propose two new metrics:

\textbf{Operation-to-configuration intensity $\IOC$} of an algorithm on a certain accelerator is the amount of accelerator operations that can be executed per byte of configuration in ops/byte.

\textbf{Configuration Bandwidth ${BW}_\text{Config}$} from a host to an accelerator is the amount of configuration bytes that the host system can set per unit time (cycles or seconds).

For a system with concurrent configuration capabilities, slotting in ${BW}_\text{Config}$ and $\IOC$ yields the following expression for attainable performance:

\begin{equation}
{P}_{\text{Attainable, Conc.}} = \min{
\begin{cases}
{P}_{\text{Peak}} \\
{BW}_{\text{Config}} \times \IOC
\end{cases}}
\label{eq:adapted_roofline}
\end{equation}

We say that performance of an algorithm on a given accelerator system is \textit{configuration bound}, if in \autoref{eq:adapted_roofline} the minimizing element stems from the configuration-related term, and compute bound otherwise.
We visualize this in dark red in \autoref{fig:adapted_roofline_model}. On this plot, a configuration bound workload (B) is located under the slanted line of the roofline, similar to a memory bound workloads in \autoref{fig:regular_roofline_model}.

Workloads in the configuration bound region hit the configuration wall: increasing the accelerator's peak performance does not increase system performance.
Here, configuration already takes more time than the accelerator's calculation itself, so performance can only be increased if ${BW}_{\text{Config}}$ or $\IOC$ are increased.
Increasing ${BW}_{\text{Config}}$ moves the slanted part of the roofline to the left, also moving the knee point and thus the boundary of the compute bound region to the left.
On the other hand, increasing $\IOC$ moves the performance point itself to the right, away from the configuration bound region.

\autoref{fig:adapted_roofline_model} also shows algorithms in the knee point (C).
These algorithms are configuring the maximal amount of settings without affecting system performance, as they spend equal time on configuration and computation.

\subsection{Sequential Accelerator Configuration Roofline}

The roofline in \autoref{eq:adapted_roofline} assumes concurrent configuration, but not all systems can do that (cfr. \autoref{subsec:seqvsconc}).
For systems that can only configure sequentially the total attainable performance has to be computed as the inverse of the attainable execution time $T_{\text{Attainable}}$:

\begin{equation}
 {P}_{\text{Attainable, Seq.}} = \frac{1}{T_{\text{Attainable}}} = \frac{1}{\frac{1}{{P}_{\text{Peak}}} + \frac{1}{{BW}_{\text{Conf.}} \times \IOC}}
\label{eq:adapted_roofline_no_overlap}
\end{equation}

This curve (shown in light red on \autoref{fig:adapted_roofline_model}) is not straight and asymptotically approaches \autoref{eq:adapted_roofline}, suggesting that losing performance on configuration cycles is unavoidable in sequential configuration systems.
Here, just pushing the algorithm to the compute bound region is not enough;
performance is dependent on how far the algorithm is from the knee point (as seen in point A on \autoref{fig:adapted_roofline_model}).

Analysis of the difference between \autoref{eq:adapted_roofline} and \ref{eq:adapted_roofline_no_overlap} tells us that the biggest performance discrepancy between concurrent and sequential configuration systems is when a system is operating in the knee point, where the system spends half of the time configuring the accelerator and half the time executing a workload.

\subsection{Effective Configuration Bandwidth} \label{subsec:effective_config_bandwidth}

The models of previous sections allow to calculate the maximum system performance if all configuration data is known and available ahead-of-time, but this is often not the case.

Often configuration data is dependent on values that are only known at runtime, such as loop-carried variables, input-dependent settings, or settings that store runtime-allocated pointers to data.
This can incur even more configuration overhead,
since the host might be required to perform extra calculations on those values 
before it can even start configuring the accelerator's registers.
A typical example of this is the bit-packing of runtime values (also shown in \autoref{fig:gemmini_setup}).

To take into account extra configuration parameter calculation, we introduce the concept of \textit{effective configuration bandwidth}.
This can be calculated by not just dividing the amount of configuration bytes $N_{\text{config bytes}}$ by the time it takes to set the registers $T_{\text{set bytes}}$, but also by the time it takes to calculate the configuration bytes $T_{\text{calc. bytes}}$ for this specific algorithm as follows:
\begin{equation}
    {BW}_{\text{Config, Eff.}} = \frac{N_{\text{config bytes}}}{T_{\text{calc. bytes}} + T_{\text{set bytes}}}
\label{eq:effective_config_bandwidth}
\end{equation}

Using ${BW}_{\text{Config, Eff.}}$, the configuration overhead estimation may be closer to reality, depending on the workload, accelerator and CPU host, as we will show in \autoref{sec:gemmini_example_numbers}.

\subsection{Relationship with the Processor Roofline} \label{subsec:3droofline}

The configuration-oriented roofline model we introduce is complementary to the classical memory-oriented processor roofline model from \cite{rooflinemodel}. A combined holistic model can be obtained by combining \autoref{eq:regular_roofline} and \autoref{eq:adapted_roofline} or \ref{eq:adapted_roofline_no_overlap}.

\begin{figure}
    \centering
    \includegraphics[width=\linewidth]{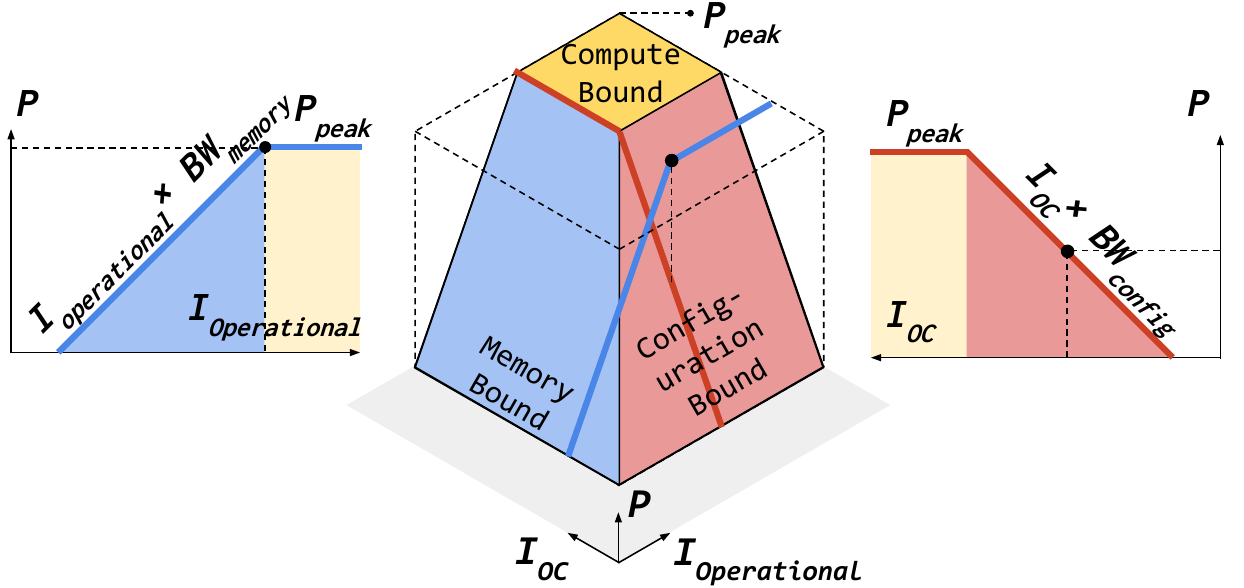}
    \caption{The processor and configuration roofline can be combined into a \textit{``roofsurface''}, that indicates whether an accelerator's performance is limited by configuration, peak memory bandwidth, or peak performance.}
    \label{fig:roofsurface}
\end{figure}

The maximum attainable performance $P_{\text{Attainable}}$ of a concurrently configured accelerator with peak performance $P_{\text{Peak}}$, peak memory bandwidth ${BW}_{\text{Memory}}$ and configuration bandwidth ${BW}_{\text{Config}}$ for a given algorithm with operational intensity ${I}_\text{Operational}$ and  operation-to-configuration intensity $\IOC$ is therefore given by:
\begin{equation}
P_{\text{Attainable}} = \min{
\begin{cases}
P_{\text{Peak}} \\
{BW}_{\text{Memory}} \times I_{\text{Operational}} \\
{BW}_{\text{Config}} \times \IOC
\end{cases}}
\label{eq:joined_roofline}
\end{equation}
This equation can be visualized as a 3-dimensional \textit{``roofsurface''} with $P_{\text{Attainable}}$ on the z-, $\IOC$ on the y-, and $I_{\text{Operational}}$ on the x-axis respectively, as seen in \autoref{fig:roofsurface}.
This figure demonstrates how an accelerator's bandwidth and compute can be perfectly balanced in the processor roofline, while its ultimate performance can still be bound by configuration in the configuration roofline.

\subsection{Example: Configuration Roofline for Gemmini}
\label{sec:gemmini_example_numbers}

As an example, let's see how Gemmini's~\cite{gemmini} output stationary matrix multiplication kernel performs on this roofline model.
In our example, this kernel calculates $\mathbf{C} = \mathbf{A}\mathbf{B} \in \mathbb{R}^{64 \times 64}, \mathbf{A} \in \mathbb{R}^{64 \times 64}, \mathbf{B} \in \mathbb{R}^{64 \times 64}$ , for a total of \textbf{524,288 ops}.

Gemmini features a 16-by-16 element systolic array, in which each processing element can perform one multiply and one accumulate operation per clock cycle, hence ${P}_{\text{Peak}} = 16\cdot16\cdot2 = \mathbf{512}$ \textbf{ops/cycle}.

To define ${BW}_{\text{Config}}$, we look into the Gemmini hardware:
The custom RoCC instructions of the 64-bit Rocket host~\cite{rocket} can configure 16 bytes at a time. 
However, since RISC-V is a load-store/architecture, it requires at least 2 instructions to put those 16 bytes into these registers.
To convert instruction counts to cycles we use 3 cycles per instruction for this host as an approximation\footnote{\label{ipcfootnote}The inverse of the harmonic mean of all instruction per cycle scores in~\cite{open_source_benchmarks}},
this yields an estimated configuration bandwidth of $\frac{16}{3\cdot3} = \mathbf{{BW}_{\text{\textbf{Config}}} \approx 1.77}$ \textbf{bytes/cycle}.

By inspecting execution traces of the spike simulator~\cite{spike}, we see that Gemmini requires 160 RoCC instrucions to configure its accelerator.
As each instruction sets 16 bytes, we can calculate the operation-to-configuration intensity as $\frac{525,288}{160 \cdot 16} = \mathbf{\IOC \approx 205.19\text{ \textbf{ops/byte}}}$.

Putting these quantities in \autoref{eq:adapted_roofline_no_overlap} --- Gemmini has sequential issuing capability --- tells us the attainable peak performance of this algorithm is only \textbf{\textit{running at 41.49\% of peak performance}}.

The previous calculation does not account for extra loop and bitwise operations to be performed by the Rocket core to pack multiple configuration bits together.
If we also trace the instructions for the bit-packing, we see that we need 935 total instructions to configure Gemmini (160 setup + 775 parameter calculation).
Again using 3 cycles per instruction\footref{ipcfootnote}, we get $\frac{160 \cdot 16}{935 \cdot 3} = \mathbf{{BW}_\text{\textbf{Config, Eff.}}  \approx 0.913 \text{\textbf{ bytes/cycle}}} $.
If we use \autoref{eq:adapted_roofline_no_overlap} with the effective instead of the theoretical configuration bandwidth, we see a \textbf{maximum attainable utilization of only 26.78\%}.

This example shows how an accelerator system can hit the configuration wall, greatly limiting achievable performance unaccounted for in the traditional processor roofline model.

\subsection{Applying the Roofline Model}\label{sec:predicting_optimisation_impact}

We now have the tools to explore how the two optimizations introduced in \autoref{sec:intro_optimisations} affect accelerator performance. We can use the roofline model to predict how observed performance will change when each optimization is applied:

\textbf{Redundant Setup Elimination} decreases the number of configuration parameters of the algorithm without affecting the actual computation. We thus expect it to \emph{increase} the operation-to-configuration intensity $\IOC$, moving the algorithm to the right of the roofline plot in \autoref{fig:adapted_roofline_model}, possibly out of the configuration-bound regime. Simulataneously, the reduced time spent configuring will result in an \emph{increase} in performance (measured in ops/cycle), moving our measurement \emph{up} in the plot, closer to peak throughput $P_{peak}$. This optimization will be most effective on code in the configuration-bound region, as it pushes code outside this regime.

\textbf{Configuration-Computation Overlap} neither changes the number of operations, nor the amount of setup of a given program. We therefore expect the observed $\IOC$ to stay constant, meaning our algorithm will \emph{not move} along the horizontal axis. Still, program duration will decrease, leading to an \emph{increase} in measured performance, moving our measurement \emph{up} in the plot. We expect this increase to be the highest in the knee-point, and proportional to the difference between the sequential- and concurrent rooflines (light and dark red lines in \autoref{fig:adapted_roofline_model}), as we are no longer limited by the sequential model and can make full use of pipelining.

\section{Implementation} \label{sec:implementation}

As we've shown, accelerators can hit the configuration wall which limits achievable performance, but the effect of this wall can be reduced by applying optimizations to configuration sequences. Current compilers however cannot perform these optimizations themselves, as they lack a way of expressing accelerator effects.

We effectively solve this problem by proposing and implementing a compiler abstraction in the form of an \ac{IR} to make the compiler aware of common accelerator configuration-related code (\autoref{sec:configuration_introduction}). We then use this abstraction to propose and implement rewrites on this IR to perform the optimizations outlined in \autoref{sec:intro_optimisations} and \autoref{sec:predicting_optimisation_impact}, and showcased in \autoref{fig:optimization_example_block_diagram}.

\subsection{Compiler Abstraction Design}
\label{sec:compiler_abstraction}

We present an approach based on the compiler framework MLIR~\cite{MLIR} and implemented in xDSL~\cite{xdsl}, a python test bed for MLIR abstractions. MLIR provides a lot of supporting infrastructure and its modular, dialect-based design lends itself to experimental extensions. MLIR's concept of a dialect provides a way to bundle operations into a clearly segmented namespace, which allows us to develop our solution as an add-on to existing compiler infrastructure, in turn allowing us to re-use a lot of the existing infrastructure.

Our dialect, called \texttt{accfg} encapsulates the configuration, launch, await programming model of accelerators:

\begin{figure}
    \centering
    \includegraphics[width=\linewidth]{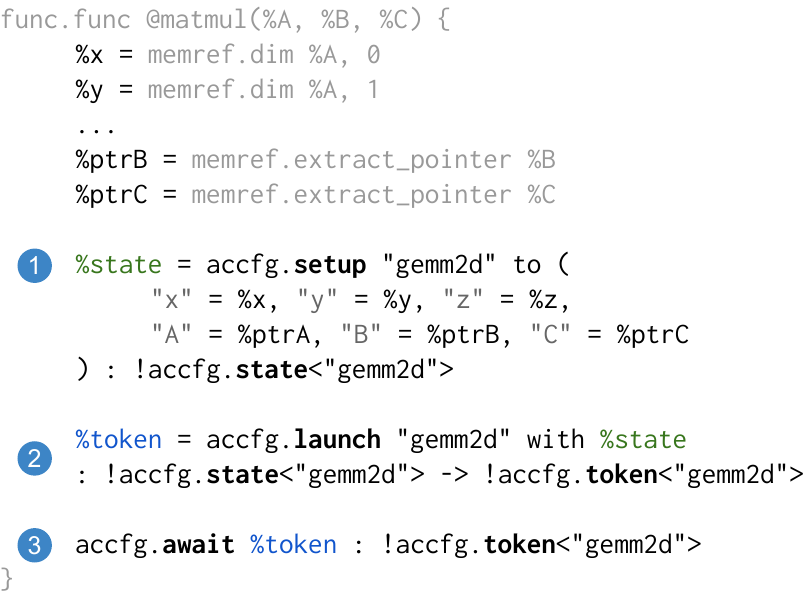}
    \caption{Simplified IR that shows how we capture common accelerator operations with our \texttt{accfg} dialect, representing accelerator configuration, launch and await operations. We highlight the setup state variable in green and the launch token variable in blue.}
    \label{fig:accfg_dialect_sample}
\end{figure}

\begin{description}
    \item[\texttt{accfg.setup}] represents a write to configuration registers without having any other effects. This produces an SSA value of type \texttt{!accfg.state}, representing the state of the accelerators' configuration file after the values have been written (\autoref{fig:accfg_dialect_sample} (1)).
     \item[\texttt{accfg.launch}] reads from a \texttt{state} variable and launches the configuration. The state variable represents (\autoref{fig:accfg_dialect_sample} (2)) the state of the accelerators' configuration registers at launch. It can optionally contain launch-semantic configurations and produces an SSA value of type \texttt{token} that can be awaited.
    \item[\texttt{accfg.await}] blocks until the accelerator completes the computation represent by the provided token. It is a no-op for sequential execution accelerators (\autoref{fig:accfg_dialect_sample} (3)).
\end{description}

Since we assume that only operations of the \texttt{accfg} dialect may modify the accelerator state, an escape hatch may be needed for stopping optimizations from crossing certain boundaries. For this, we provide an attribute that can be added to any non-\texttt{accfg} operation to signal that it destroys the accelerator state: \texttt{\#accfg.effects<all>}.

Conversely, function calls to external functions cannot be inspected by the compiler, and therefore pose optimization barriers. To stop a \texttt{printf} call from disabling all optimizations, we provide another version of this attribute to signal that state is preserved across a specific operation: \texttt{\#accfg.effects<none>}. A more fine-grained effects interface for foreign operations was not needed for this work.

We place an additional constraint on the IR: only one \texttt{state} variable may be ``live'' at any point in time per accelerator: A \texttt{setup} produces a state variable, from which can be read by \texttt{launch} operations until the next ``state'' is produced by the next \texttt{setup} operation. A \texttt{setup} operation can receive an input state, allowing it to determine a ``setup delta'' between itself and the previous state. This represents the core idea of our abstraction.

This IR intends to capture the semantics of inline-assembly instructions like the ones used to configure OpenGeMM and Gemmini. Our model does not address failed configuration writes or accelerator faults, which is assumed to happen at a higher level and is thus not part of this work (e.g., system-level trap handlers). We note that neither OpenGeMM nor Gemmini have features for tracking such faults.

\subsection{Compilation Flow}\label{sec:accfg-passes}

\begin{figure}
    \centering
    \includegraphics[width=\linewidth]{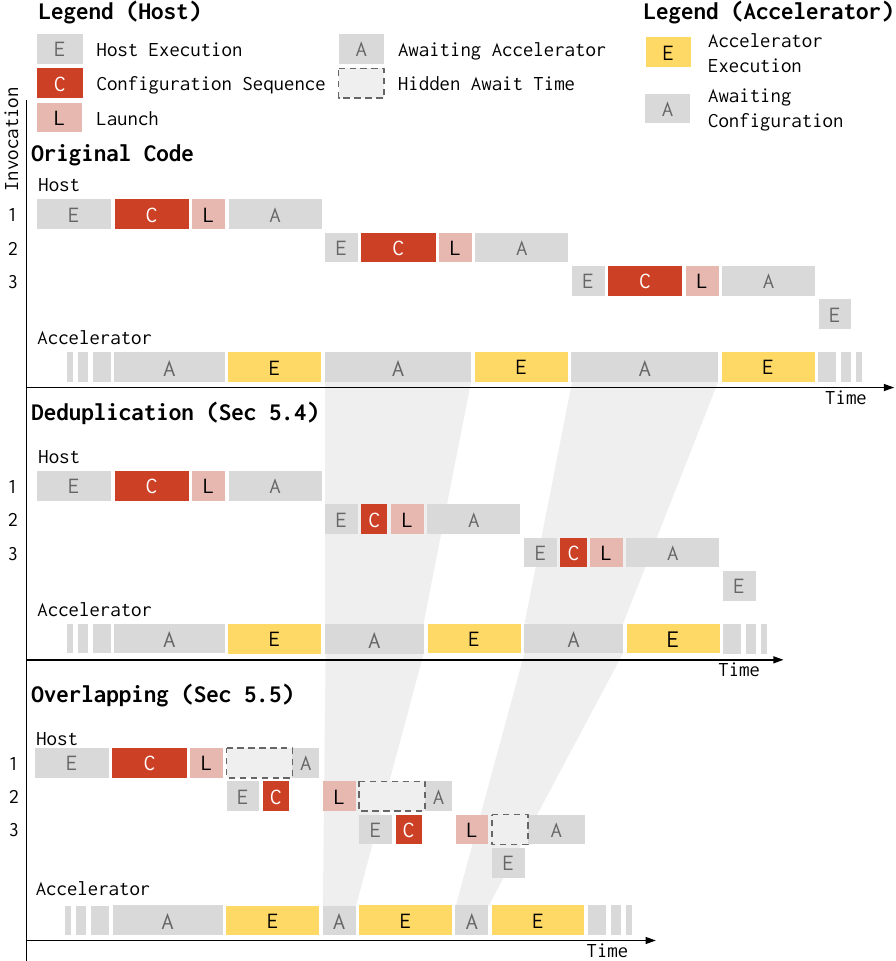}
    \caption{Our optimizations drastically reduce accelerator idle time by first removing duplicate setups from repeated configuration invocations (\autoref{subsec:configuration_deduplication}) and then perform pipelining to overlap host compute and configuration calls with accelerator execution (\autoref{subsec:configuration_overlap_pass}).}
    \label{fig:optimization_example_block_diagram}
\end{figure}

\begin{figure}
    \centering
    \includegraphics[width=\linewidth]{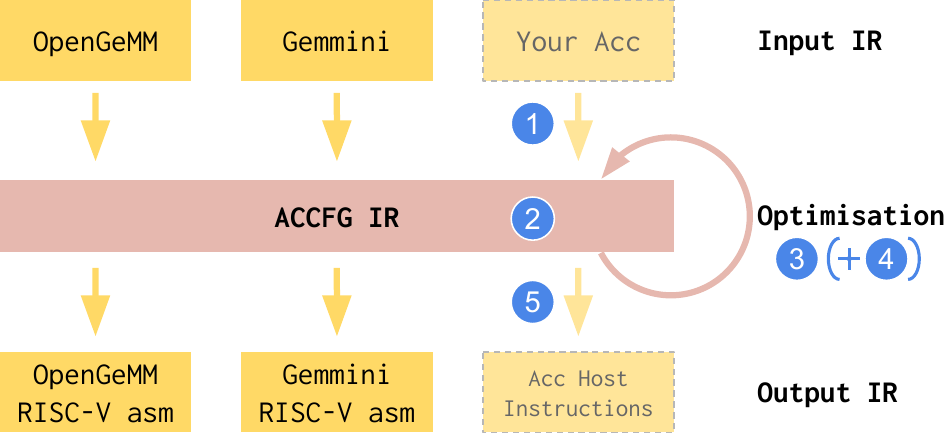}
    \caption{Our dialect design allows it to be inserted into multiple lowering pipelines. All target architectures must implement their own translations (steps 1 and 5), but use the same \texttt{accfg} abstractions, and share a large part of the available rewrites (steps 2, 3 and 4).}
    \label{fig:compilation_flow}
\end{figure}

Adding \texttt{accfg} to the compilation flow requires the introduction of five additional steps (highlighted as 1 to 5 in \autoref{fig:compilation_flow}). In step 1 the compiler must represent accelerator dispatches through disjunct setup/launch/await clusters in the IR similar to \autoref{fig:accfg_dialect_sample}. In step 2, we establish a connection between these clusters by connecting the accelerator configuration \texttt{state} variable of these separate clusters automatically. In step 3 we optimize the IR by performing redundant store elimination. Step 4 optionally performs setup overlap (can only be applied for concurrent configuration accelerator systems). Finally step 5 is needed to convert the \texttt{accfg} operations into the actual setup sequences for the target architecture.

The dialect conversions (step 1 and 5) needed at the start and end of our optimization passes represent the only accelerator specific transformations in this pipeline. Every supported accelerator will need their own set of translation passes here, but can re-use all other passes provided in this work.
These translations represent simple search-and-replace operations on the \ac{IR}, for which MLIR already provides a powerful interface. We will therefore not go into further detail on them.

Due to the \texttt{accfg} operations properly defining their effects instead of being opaque \texttt{volatile} blocks, we immediately benefit from MLIRs already implemented optimizations for all major modern CPU architectures such as more aggressive constant folding, common-subexpression-elimination and loop-invariant-code-motion. This is already effective in reducing the effective configurational intensity.

The remaining steps (2, 3 and 4) represent the core ideas presented in this paper, and are outlined below.

\subsection{State Tracing} \label{subsec:state_tracing}
Reasoning about sequences of setup operations requires us to establish an order of invocations. We do this by introducing a state-variable for each accelerator. With this, we can model read/write effects of the accelerators' configuration registers with the tools provided within MLIR. This is inspired by the concept of memory SSA \cite{memssa}, which encodes clobbering and alias analysis directly into the def-use chain.

We inspect for each \texttt{setup} operation whether we can determine the previous live \texttt{state} variable through static analysis of control flow. The IR is then rewritten to connect these states by adding the prior \texttt{state} as an input to the setup operation. Our implementation does this for straight-line code, \texttt{if}/\texttt{else} branches, and \texttt{for}-loops, making pessimistic assumptions about invisible code (such as function calls).

We ensure correctness by assuming all unknown operations and function calls could read/write the state unless annotated with an \texttt{\#accfg.effects<none>} attribute.

\subsection{Configuration Deduplication} \label{subsec:configuration_deduplication}
We now have all the necessary information expressed in the IR needed to perform configuration deduplication. Due to the form of state-traced \ac{IR}, we are able to infer which register holds which value. 

We can now visit every \texttt{setup} operation and inspect the operation that produced the prior input state, to construct the state of the accelerators' configuration registers. On a high level, it walks the use-def chain backwards to discover all setup operations that could have clobbered the configuration and constructs a map of known fields and their values.

We use this information to then remove setup instructions that represent a redundant write of a value to a setup field, i.e. a write of the same value to the same location.
We check for SSA-Value equivalence as a proxy for value-equivalence. Due to the properties of SSA, we don't actually need to infer the runtime-value at compile time, and instead rely on the fact that SSA values can't be re-assigned during the course of a program, meaning that the same SSA-value will always contain the same value at runtime.

MLIR's powerful common sub-expression elimination, loop-invariant code motion (LICM) and canonicalization passes go through great lengths to ensure that SSA variables referencing the same value are deduplicated. Therefore, we are able to leverage SSA-level equivalence to perform deduplication. Even though these optimizations are also applied in regular compilers, they can never perform them on external register state, something our compiler explicitly enables.

\subsubsection{Optimizing through Control Flow}

\begin{figure*}
    \centering
    \includegraphics[width=\linewidth]{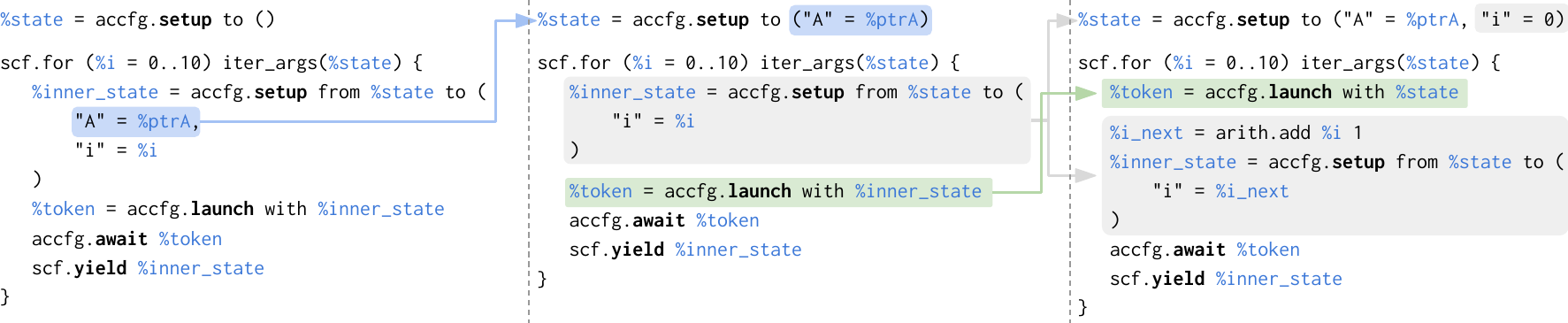}
    \caption{Visualization of our loop-level optimizations on simplified IR, showing loop-invariant code motion (blue) in the first transition and overlap (gray-green) in the second. Types and accelerator names have been omitted.}
    \label{fig:loop_motion_and_overlap}
\end{figure*}

To increase the potency of the configuration deduplication, we perform two additional rewrites right before attempting deduplication: Lifting setup calls into branching logic, and lifting loop-invariant setup fields out of loops. Hoisting into branches is motivated by the fact that our inference has to take the intersection of the two sides of the branch when the control flow forks, removing possible optimization potential as we can't know after the branch which path was executed. By hoisting the setup into the branch, we re-gain a linear chain of setup calls creating more opportunities for deduplication.

Our loop-hoisting logic closely follows MLIRs existing LICM pass, with the additional constraint that we can only hoist parameters that stay constant throughout the whole loop body (e.g. if there are two consecutive launches with different parameters inside the loop, we can't hoist them even though they are technically loop-invariant). The pass inspects each setup parameter separately to discern if it is loop-dependent or not, and moves it to a newly created setup operation right in front of the loop if it is determined that the value can be moved safely (See second block of \autoref{fig:loop_motion_and_overlap} and \autoref{fig:optimization_example_block_diagram}). After these two rewrites are performed, we run two more clean-up passes that remove empty \texttt{setup} operations, and merge \texttt{setup} operations which have no launch statements in between.

This optimization pass ultimately leaves us with a functionally equivalent program that needs to spend much less time on configuring registers.

\subsection{Configuration Overlap} \label{subsec:configuration_overlap_pass}
In order to fully exploit the concurrency in many modern accelerator setups, we also implement the second optimization that schedules code to run \emph{concurrently} with the accelerator. We do this by pushing suitable code up \emph{before} the previous \texttt{await} statement. While this optimization is relatively easy to implement for straight-line code, overlapping setup and computation inside loops is difficult, as a setup sequence \emph{must} happen before the accelerator is launched.

As part of this optimization, we implement a form of software-pipelining that allows us to transform looping code into IR
suitable for overlapping configuration and computation. This is done by identifying suitable 
setup sequences (more on this further down in this section), and pushing them ``up'' one loop iteration. This is done in two steps:

\begin{enumerate}
    \item First we insert a copy of the setup operation before the loop starts, replacing all uses of the loop counter with the lower bound for the loop.
    \item Then we change the setup sequence inside the loop to use an incremented loop counter, making sure to leave other uses of the loop counter unaffected.
\end{enumerate}

As stated earlier, not all setup sequences can be rearranged like this. Our setup sequence identification algorithm checks whether all operations are \emph{pure},
meaning that no read-before-write errors are introduced when they are moved around. If the setup sequence contains impure operations that cannot be moved safely, a partial move of the setup operation could still be performed, although this is not implemented in our current infrastructure.

After the loops have been rewritten to a suitable form for overlapping, a relatively simple block-level rewrite pattern can overlap configuration and accelerator execution on straight-line code by looking for a setup operation with an input state that is launched and awaited earlier inside the same block. If we are able to identify such a setup, we move it up and in front of the identified \texttt{await} operation\footnote{We move the entire \emph{setup sequence}, consisting of both the setup and pure (non-effecting) operations that calculate its inputs.}. This is shown in the third block of \autoref{fig:loop_motion_and_overlap} and \autoref{fig:optimization_example_block_diagram}.

\section{Evaluation}
\label{sec:evaluation}

To evaluate our proposed abstraction and optimization passes,
we demonstrate in this section how much we can reduce the configuration overhead and increase performance of tiled matrix multiplication benchmarks.
We consider two targets, Gemmini~\cite{gemmini}, which is a sequential-configuration architecture (so configuration-overlap is not possible) and OpenGeMM~\cite{opengemm}
which is a concurrent-configuration architecture.
While these platforms are both optimized for matrix multiplications,
their host CPU and software stacks are very different --- this section sets out to show that
\texttt{accfg} is general enough to be leveraged for a variety of architectures.
In addition to these case studies, we will look at how the performance results fit within our roofline model (\autoref{sec:roofline_results}).

\subsection{Tiled Matrix Multiplication on Gemmini} \label{subsec:performance_gemmini}

In order to approximate the configuration overhead on Gemmini~\cite{gemmini},
we implement \texttt{accfg} lowerings to run Gemmini's bare-metal weight-stationary tiled matrix multiply experiments on various matrix sizes in the instruction-accurate spike simulator~\cite{spike}.
We trace the number of configuration bytes and setup and calculation instructions with Rocket's performance counters in spike to calculate ${BW}_{\text{Config, Eff.}}$, and $\IOC$.
We use $P_{\text{Peak}}= 512$ ops/cycle and an average of 3 cycles per instruction, extracted from~\cite{open_source_benchmarks} for the Rocket core and use \autoref{eq:adapted_roofline_no_overlap} as a proxy for the attainable performance.

\begin{figure}
    \centering
    \includegraphics[width=\linewidth]{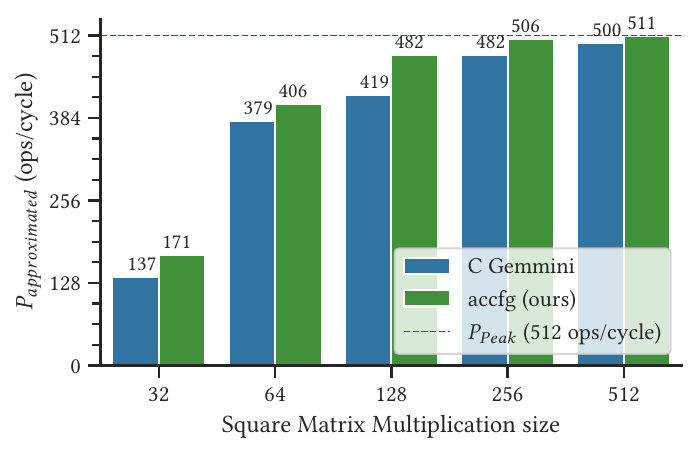}
    \caption{
    The attainable performance of Gemmini's weight-stationary matrix multiplication mode on various matrix sizes
    shows that accfg enables an overall geomean performance uplift of 11\%.
    }
    \label{fig:gemmini_performance}
\end{figure}

As a baseline we use Gemmini's experiments in C code, compiled with GCC 13.2.0 and optimization level \texttt{-O2}, which is the optimization level used in their experiments' repository.
For our optimized binaries we progressively lower a matrix multiplication in MLIR using \texttt{accfg} to LLVM IR with LLVM inline assembly for the RoCC instructions.
The resulting LLVM IR is compiled by clang with \texttt{-O2} and \texttt{--target=}\\\texttt{generic-riscv64}, and linked to an adapted C main file with GCC, using the same settings as the baseline.
The resulting attainable performance is shown in \autoref{fig:gemmini_performance}, where we see a geomean performance improvement of 11\%.

The experimental results show an uplift in Gemmini's performance.
Gemmini's weight-stationary kernel does not have a lot of room for improvement: it sets up a lot less parameters than its output-stationary counterpart, and does not allow for configuration-computation overlap.
Nonetheless we still see performance improvements due to better constant
folding and loop unrolling. Our \texttt{accfg} dialect
enables these improvements because it provides more accurate
semantics to the compiler than volatile assembly statements.
In Gemmini's output stationary flow (which we do not evaluate
here), we would expect to see larger performance improvements.

Our optimizations appear to have the biggest impact at a size of 128. Smaller sizes only require a single invocation, preventing us from benefiting from deduplication, while larger sizes are heavily compute bound, negating any benefits gained from deduplication. Still, at a size of 128 we see a speed-up of $\approx 15\%$ over the baseline produced by GCC\@.  Overall,
we see performance improvements averaging $11\%$.

\subsection{Tiled Matrix Multiplication on OpenGeMM} \label{subsec:perf_concurrent}

\begin{figure}
    \centering
    \includegraphics[width=\linewidth]{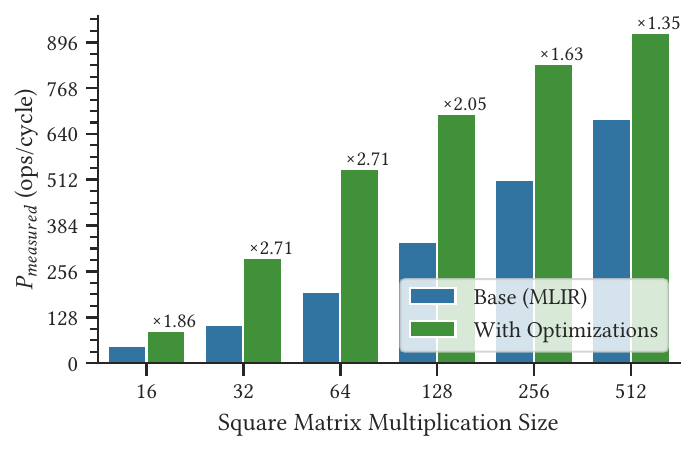}
    \caption{
    The performance of matrix multiplications on OpenGeMM~\cite{opengemm} with various matrix sizes.
    We can see that the optimizations that accfg enables
    unlock significant extra performance, with
    a geomean of improvement of 2x.
    }
    \label{fig:snax_measured_perf}
\end{figure}

We also run our matrices on OpenGeMM~\cite{opengemm}, a concurrent configuration matrix multiplication accelerator system with peak performance of 1024 ops/cycle and a tiny in-order RISC-V core \cite{snitch}.
We run tiled matrix multiplications of K-by-K squared matrices, using a tile size of 8-by-K-by-8.  
inside a cycle-accurate model through Verilator.
To properly gauge the configuration overhead, we only measure the performance of the tiling loop itself and turn off all memory copies between internal scratchpads. 
All binaries are compiled with our \texttt{accfg}-based MLIR flow, but the base results do not perform any configuration deduplication or configuration overlap.
\autoref{fig:snax_measured_perf} shows the comparison of these results. Our optimizations help us achieve a significant geomean speed-up of 2x.

\begin{figure}
    \centering
    \setstackgap{S}{0pt} 

    \centerline{\stackinset{c}{}{c}{}{\includegraphics[width=\linewidth]{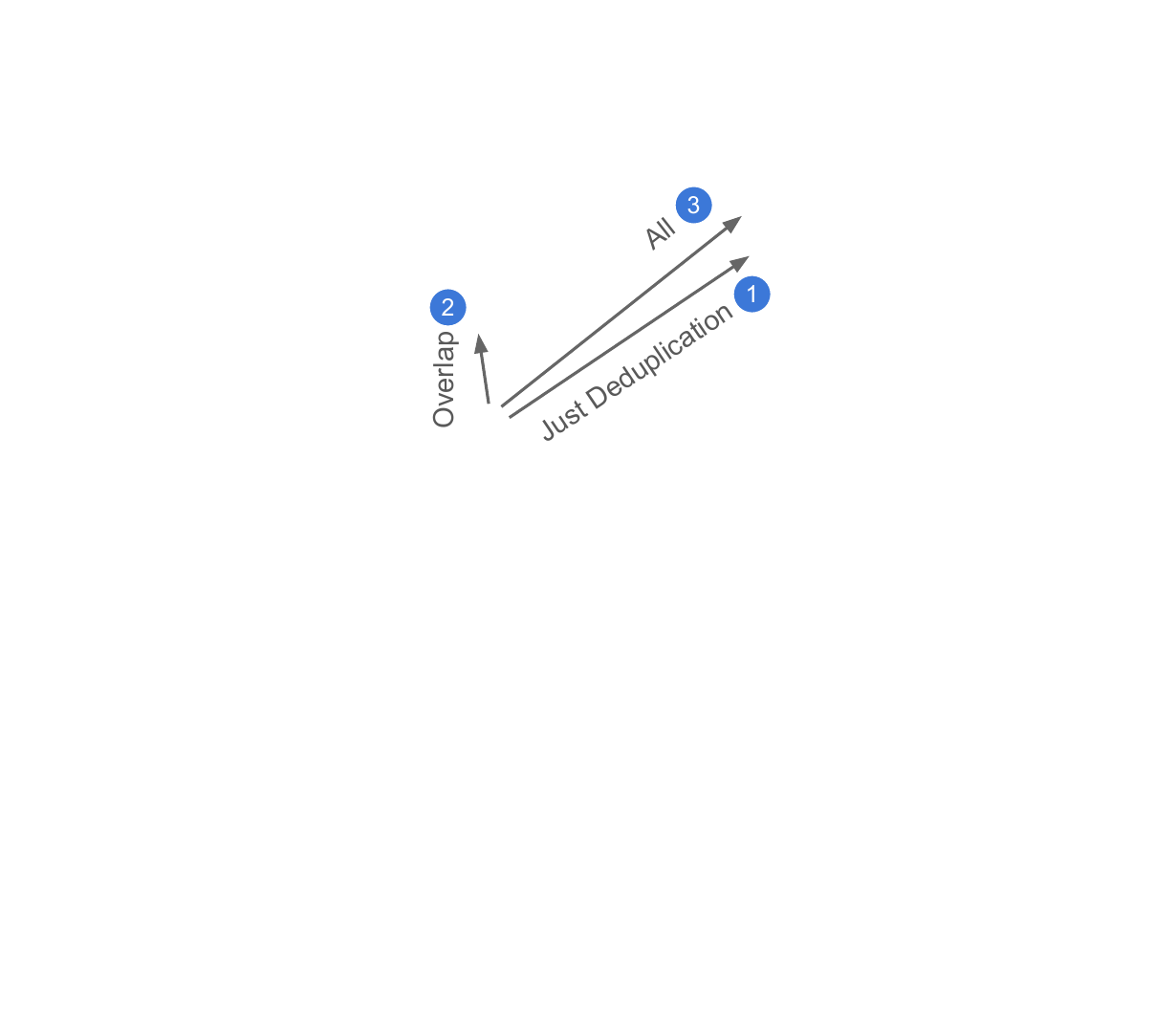}}%
                              {\includegraphics[width=\linewidth]{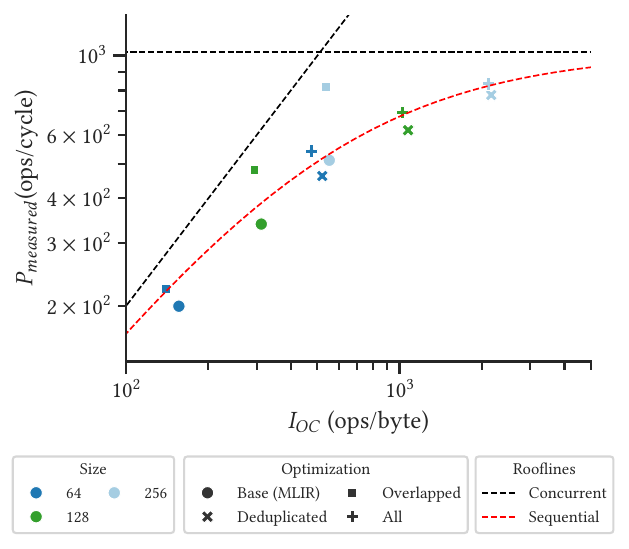}}}

    \caption{Plotting our measurements together with both the sequential (red) and concurrent configuration (black) rooflines of OpenGeMM~\cite{opengemm} shows the power of our optimizations and model. Our overlap optimization lifts the peak performance of the accelerator. Our deduplication optimization increases $\IOC$ and lifts peak performance}
    \label{fig:snax_roofline_plot}
\end{figure}

\subsubsection{Comparing Results with Roofline Model}
\label{sec:roofline_results}

We also plot our results for OpenGeMM into its roofline chart in \autoref{fig:snax_roofline_plot}. We can compare the impact of individual optimizations against the intuition developed in \autoref{sec:predicting_optimisation_impact}. As outlined in that section, we can see that removing redundant setups (deduplication), moves the measurement point up and to the right in the plot (arrow 1), signifying an increase in throughput and a reduction of configuration overhead. For size 128, we can clearly see that performing the optimization moves the algorithm \emph{out of the configuration-bound regime}.

Introducing overlap moves the measurement point \emph{up} as predicted (arrow 2), and the distance moved correlates directly with the difference between the sequential- and concurrent rooflines. Performing both deduplication and overlapping yields the biggest speed-ups (arrow 3).

In general, we can conclude that our optimizations effectively enable moving accelerators away from the configuration bound region towards the compute bound region.

\section{Related Work}
\label{sec:related_work}

\textbf{Performance modeling for configuration overhead.}
As mentioned in \autoref{subsec:the_configuration_roofline}, we build our configuration roofline on work from Williams et. al. \cite{rooflinemodel}, which discusses how to assess performance bounds related to datapath and memory subsystems. From the many extensions that additionally inspired us to create the configuration roofline \cite{cache_roofline,distributed_roofline,gpumicrobenchmark,amdgpu,nvidiagpu}, we highlight two works\cite{amdgpu, nvidiagpu} which discuss an instruction roofline tailored for predicting performance in GPU datapaths, furthermore we highlight the work of Cardwell et al. \cite{distributed_roofline} which adds an additional dimension to the roofline for communication overhead in HPC systems. We believe these models are orthogonal to our work in the context of more complex systems.
Other work \cite{laa_performance_overhead} also discusses the adverse effect of slow host-accelerator communication overhead, however the authors do not differentiate between data transfer and configuration, which our model separates. 
The LogCA~\cite{logca} model determines the potential speed-up from offloading work onto an accelerator, how design decisions affect this, and how to change the design to maximize speed-up, but does not model the configuration trade-off we discuss.

\noindent\textbf{Accelerators suffering from configuration overhead.}
We find many works that evidence accelerators suffering from the configuration wall, and which propose ad-hoc hardware/software solutions for getting their accelerators out of the configuration-bound region. 
Cheshmikhani et. al. \cite{riscv_multi_accelerator_control} propose custom RISC-V instructions set architecture (ISA) extension for concurrent configuration accelerators in a multi-process setting.
They report up to 10x speed-up by switching from an memory-mapped input/output (MMIO) driver-based solution to an ISA extension.
Work in the field of cryptographic accelerators~\cite{crypto_interface,crypto_facv} also evidence a speed-up obtained by moving from an MMIO, to and ISA-based solution for the same accelerator.
Optimus Prime~\cite{rpc_optimus_prime} explicitly mentions that kernel driver bypass is crucial for low-latency data transformation acceleration in remote procedure calls (RPCs) for cloud computing workloads.
Kim et. al. \cite{cgra_mapping_optimization} observe a performance dip in host-accelerator communication for coarse-grained reconfigurable arrays (CGRAs) which they resolve with extra hardware capabilities and a loop scheduling trick.
Aurora \cite{aurora} proposes a software runtime system and a hardware configuration interface to share accelerators among multiple processes with a low reconfiguration overhead.
Mozart \cite{mozart} explicitly mentions \emph{data} and \emph{control taxes} which correspond to our notions of memory-bound and configuration-bound regions in the roofline model.
To tackle the configuration-overhead problem, they design the Accelerator Synchronization Interface (ASI), which can synchronize accelerators through a programming interface attached to coherent shared-memory.
For each of these approaches our configuration roofline can be used to quantify the configuration overhead.
We also believe that most bare-metal accelerator approaches can benefit from our MLIR dialect in their compilation pipeline.

\noindent\textbf{MLIR-based compilers for specialized accelerators.}
We see a lot of similarities with AXI4MLIR~\cite{axi4mlir}, which automates host-to-accelerator code generation.
The authors implement and suggest some ad-hoc configuration-related optimizations but do not quantify the configuration overhead specifically.
Rather, the authors explore different tiling strategies for optimizing data movement, which is orthogonal to our work. 
Similar to our configuration overlap pass, \texttt{tpp-mlir}, described by Golin et. al. \cite{intel_tpp_mlir}, hoists AMX tile configuration instruction out of the for loop in MLIR.
Our approach works very similar, but is platform-agnostic.
For Gemmini, an MLIR-based project is available through the Buddy compiler~\cite{buddycompiler}.
In this framework, an MLIR-dialect specific to Gemmini is available, but the dialect is not capable of doing configuration overhead elimination.

\section{Conclusions and Outlook}
\label{sec:conclusion_outlook}

In this work we show how modern accelerator implementations are fundamentally limited by a \emph{configuration wall}. We characterize this wall through a novel roofline model. The roofline model provides us with a framework to reason about performance limitations present in host-to-accelerator configuration interaction. We provide intuition on why current compilers cannot work around these bottlenecks effectively, and provide a compiler-based solution to solve this problem.

We evaluate our implementation on two platforms, showing performance uplifts which demonstrate the effectiveness of our optimizations in moving workloads out of the configuration bound regime. By plotting effects of individual optimizations into our roofline, we confirm intuition built through our novel model. These results confirm the real-world impact and usefulness of our work.

\onecolumn
\begin{multicols}{2}

We see a lot of potential for future work based on this approach. Exploring a more fine-grained system for declaring effects to reason about accelerator state across function call boundaries, which is not handled by the current approach. Another aspect that could improve potency of these optimizations for more complex workloads could come from better handling of conditional invocations.
We are also really interested in seeing how this work can be leveraged to optimize configuration for systems with other setup schemes, such as the use of FIFOs and queues to create software defined pipelines in Cohort\cite{cohort}.
We also are very interested to see how our work applies to multi-accelerator systems \cite{aurora, cohort, mozart}, which sometimes have decentralized configuration possibilities \cite{cohort, mozart}.

This work represents a promising step towards making accelerator configuration more approachable and optimizable. We are confident that this work already provides valuable insights and tools for both hardware vendors, compiler builders and library writers to make their systems less impacted by configuration overhead.

\begin{acks}
  This work has received funding from the European Union’s Horizon EUROPE research and innovation program under grant agreement no. \grantnum{Horizon Europe}{101070375 (CONVOLVE)} and \grantnum{Horizon Europe}{101088865 (BINGO)}, Research Foundation Flanders (FWO) under grant \grantnum{FWO}{1SE7723N}, Flanders AI Research Program (FAIR) and the \grantsponsor{UKRI}{UKRI EPSRC Centre for Doctoral Training in Machine Learning Systems}{\url{https://gtr.ukri.org/projects?ref=EP\%2FY03516X\%2F1}} project ref. \grantnum{UKRI}{EP/Y03516X/1}.
\end{acks}

\bibliographystyle{ACM-Reference-Format}
\bibliography{references}

@inproceedings{distributed_roofline,
    author = {Cardwell, David and Song, Fengguang},
    title = {An Extended Roofline Model with Communication-Awareness for Distributed-Memory HPC Systems},
    year = {2019},
    isbn = {9781450366328},
    publisher = {Association for Computing Machinery},
    address = {New York, NY, USA},
    url = {https://doi.org/10.1145/3293320.3293321},
    doi = {10.1145/3293320.3293321},
    booktitle = {Proceedings of the International Conference on High Performance Computing in Asia-Pacific Region},
    pages = {26–35},
    numpages = {10},
    location = {Guangzhou, China},
    series = {HPCAsia '19}
}

@ARTICLE{cache_roofline,
  author={Ilic, Aleksandar and Pratas, Frederico and Sousa, Leonel},
  journal={IEEE Computer Architecture Letters}, 
  title={Cache-aware Roofline model: Upgrading the loft}, 
  year={2014},
  volume={13},
  number={1},
  pages={21-24},
  doi={10.1109/L-CA.2013.6}
}

@inproceedings{tpu,
    author = {Jouppi, Norman P. and Young, Cliff and Patil, Nishant and Patterson, David and Agrawal, Gaurav and Bajwa, Raminder and Bates, Sarah and Bhatia, Suresh and Boden, Nan and Borchers, Al and Boyle, Rick and Cantin, Pierre-luc and Chao, Clifford and Clark, Chris and Coriell, Jeremy and Daley, Mike and Dau, Matt and Dean, Jeffrey and Gelb, Ben and Ghaemmaghami, Tara Vazir and Gottipati, Rajendra and Gulland, William and Hagmann, Robert and Ho, C. Richard and Hogberg, Doug and Hu, John and Hundt, Robert and Hurt, Dan and Ibarz, Julian and Jaffey, Aaron and Jaworski, Alek and Kaplan, Alexander and Khaitan, Harshit and Killebrew, Daniel and Koch, Andy and Kumar, Naveen and Lacy, Steve and Laudon, James and Law, James and Le, Diemthu and Leary, Chris and Liu, Zhuyuan and Lucke, Kyle and Lundin, Alan and MacKean, Gordon and Maggiore, Adriana and Mahony, Maire and Miller, Kieran and Nagarajan, Rahul and Narayanaswami, Ravi and Ni, Ray and Nix, Kathy and Norrie, Thomas and Omernick, Mark and Penukonda, Narayana and Phelps, Andy and Ross, Jonathan and Ross, Matt and Salek, Amir and Samadiani, Emad and Severn, Chris and Sizikov, Gregory and Snelham, Matthew and Souter, Jed and Steinberg, Dan and Swing, Andy and Tan, Mercedes and Thorson, Gregory and Tian, Bo and Toma, Horia and Tuttle, Erick and Vasudevan, Vijay and Walter, Richard and Wang, Walter and Wilcox, Eric and Yoon, Doe Hyun},
    title = {In-Datacenter Performance Analysis of a Tensor Processing Unit},
    year = {2017},
    isbn = {9781450348928},
    publisher = {Association for Computing Machinery},
    address = {New York, NY, USA},
    url = {https://doi.org/10.1145/3079856.3080246},
    doi = {10.1145/3079856.3080246},
    booktitle = {Proceedings of the 44th Annual International Symposium on Computer Architecture},
    pages = {1–12},
    numpages = {12},
    location = {Toronto, ON, Canada},
    series = {ISCA '17}
}

@article{amdgpu,
    author = {Leinhauser, Matthew and Widera, Ren\'{e} and Bastrakov, Sergei and Debus, Alexander and Bussmann, Michael and Chandrasekaran, Sunita},
    title = {Metrics and Design of an Instruction Roofline Model for AMD GPUs},
    year = {2022},
    issue_date = {March 2022},
    publisher = {Association for Computing Machinery},
    address = {New York, NY, USA},
    volume = {9},
    number = {1},
    issn = {2329-4949},
    url = {https://doi.org/10.1145/3505285},
    doi = {10.1145/3505285},
    journal = {ACM Trans. Parallel Comput.},
    month = jan,
    articleno = {1},
    numpages = {14}
}

@article{nvidiagpu,
    author = {Ding, Nan and Awan, Muaaz and Williams, Samuel},
    title = {Instruction Roofline: An insightful visual performance model for GPUs},
    journal = {Concurrency and Computation: Practice and Experience},
    volume = {34},
    number = {20},
    pages = {e6591},
    doi = {https://doi.org/10.1002/cpe.6591},
    url = {https://onlinelibrary.wiley.com/doi/abs/10.1002/cpe.6591},
    eprint = {https://onlinelibrary.wiley.com/doi/pdf/10.1002/cpe.6591},
    year = {2022}
}

@article{gpumicrobenchmark,
  title={A quantitative roofline model for GPU kernel performance estimation using micro-benchmarks and hardware metric profiling},
  author={Konstantinidis, Elias and Cotronis, Yiannis},
  journal={Journal of Parallel and Distributed Computing},
  volume={107},
  pages={37--56},
  year={2017},
  publisher={Elsevier}
}

@inproceedings{cohort,
    author = {Wei, Tianrui and Turtayeva, Nazerke and Orenes-Vera, Marcelo and Lonkar, Omkar and Balkind, Jonathan},
    title = {Cohort: Software-Oriented Acceleration for Heterogeneous SoCs},
    year = {2023},
    isbn = {9781450399180},
    publisher = {Association for Computing Machinery},
    address = {New York, NY, USA},
    url = {https://doi.org/10.1145/3582016.3582059},
    doi = {10.1145/3582016.3582059},
    booktitle = {Proceedings of the 28th ACM International Conference on Architectural Support for Programming Languages and Operating Systems, Volume 3},
    pages = {105–117},
    numpages = {13},
    location = {Vancouver, BC, Canada},
    series = {ASPLOS 2023}
}

@inproceedings{mozart,
    author = {Suresh, Vignesh and Mishra, Bakshree and Jing, Ying and Zhu, Zeran and Jin, Naiyin and Block, Charles and Mantovani, Paolo and Giri, Davide and Zuckerman, Joseph and Carloni, Luca P. and Adve, Sarita V.},
    title = {Mozart: Taming Taxes and Composing Accelerators with Shared-Memory},
    year = {2024},
    isbn = {9798400706318},
    publisher = {Association for Computing Machinery},
    address = {New York, NY, USA},
    url = {https://doi.org/10.1145/3656019.3676896},
    doi = {10.1145/3656019.3676896},
    booktitle = {Proceedings of the 2024 International Conference on Parallel Architectures and Compilation Techniques},
    pages = {183–200},
    numpages = {18},
    location = {Long Beach, CA, USA},
    series = {PACT '24}
}

@ARTICLE{riscv_multi_accelerator_control,
  author={Cheshmikhani, Elham and Peccerillo, Biagio and Mondelli, Andrea and Bartolini, Sandro},
  journal={IEEE Access},
  title={A General Framework for Accelerator Management Based on ISA Extension},
  year={2022},
  volume={10},
  number={},
  pages={120702-120713},
  doi={10.1109/ACCESS.2022.3222346}}

@misc{xdsl,
      title={Sidekick compilation with xDSL},
      author={Mathieu Fehr and Michel Weber and Christian Ulmann and Alexandre Lopoukhine and Martin Lücke and Théo Degioanni and Michel Steuwer and Tobias Grosser},
      year={2024},
      eprint={2311.07422},
      archivePrefix={arXiv},
      primaryClass={cs.PL},
      url={https://arxiv.org/abs/2311.07422},
}

@INPROCEEDINGS{opengemm,
  author={Xiaoling Yi and Ryan Antonio and Joren Dumoulin and Jiacong Sun and Josse Van Delm and Guilherme Paim and Marian Verhelst},
  booktitle={2025 30th Asia and South Pacific Design Automation Conference (ASP-DAC)},
  title={{OpenGeMM}: A High-Utilization GeMM Accelerator Generator
with Lightweight RISC-V Control and Tight Memory Coupling},
  year={2025},
  volume={},
  number={},
  pages={},
  doi={}}

@ARTICLE{snitch,
  author={Zaruba, Florian and Schuiki, Fabian and Hoefler, Torsten and Benini, Luca},
  journal={IEEE Transactions on Computers},
  title={Snitch: A Tiny Pseudo Dual-Issue Processor for Area and Energy Efficient Execution of Floating-Point Intensive Workloads},
  year={2021},
  volume={70},
  number={11},
  pages={1845-1860},
  doi={10.1109/TC.2020.3027900}}

@article{buddycompiler,
  title={Compiler Technologies in Deep Learning Co-Design: A Survey},
  author={Zhang, Hongbin and Xing, Mingjie and Wu, Yanjun and Zhao, Chen},
  journal={Intelligent Computing},
  year={2023},
  publisher={AAAS}
}

@INPROCEEDINGS {AXI4MLIR,
author = {N. Agostini and J. Haris and P. Gibson and M. Jayaweera and N. Rubin and A. Tumeo and J. L. Abellan and J. Cano and D. Kaeli},
booktitle = {2024 IEEE/ACM International Symposium on Code Generation and Optimization (CGO)},
title = {AXI4MLIR: User-Driven Automatic Host Code Generation for Custom AXI-Based Accelerators},
year = {2024},
volume = {},
issn = {},
pages = {143-157},
doi = {10.1109/CGO57630.2024.10444801},
url = {https://doi.ieeecomputersociety.org/10.1109/CGO57630.2024.10444801},
publisher = {IEEE Computer Society},
address = {Los Alamitos, CA, USA},
month = {mar}
}

@article{cgra_mapping_optimization,
author = {Kim, Yongjoo and Lee, Jongeun and Mai, Toan X. and Paek, Yunheung},
title = {Improving performance of nested loops on reconfigurable array processors},
year = {2012},
issue_date = {January 2012},
publisher = {Association for Computing Machinery},
address = {New York, NY, USA},
volume = {8},
number = {4},
issn = {1544-3566},
url = {https://doi.org/10.1145/2086696.2086711},
doi = {10.1145/2086696.2086711},
journal = {ACM Trans. Archit. Code Optim.},
month = {jan},
articleno = {32},
numpages = {23}
}

@inproceedings{open_source_benchmarks,
author = {D\"{o}rflinger, Alexander and Albers, Mark and Kleinbeck, Benedikt and Guan, Yejun and Michalik, Harald and Klink, Raphael and Blochwitz, Christopher and Nechi, Anouar and Berekovic, Mladen},
title = {A comparative survey of open-source application-class RISC-V processor implementations},
year = {2021},
isbn = {9781450384049},
publisher = {Association for Computing Machinery},
address = {New York, NY, USA},
url = {https://doi.org/10.1145/3457388.3458657},
doi = {10.1145/3457388.3458657},
booktitle = {Proceedings of the 18th ACM International Conference on Computing Frontiers},
pages = {12–20},
numpages = {9},
location = {Virtual Event, Italy},
series = {CF '21}
}

@inproceedings{mlir,
  author={Lattner, Chris and Amini, Mehdi and Bondhugula, Uday and Cohen, Albert and Davis, Andy and Pienaar, Jacques and Riddle, River and Shpeisman, Tatiana and Vasilache, Nicolas and Zinenko, Oleksandr},
  booktitle={2021 {{IEEE/ACM}} International Symposium on Code Generation and Optimization (CGO)},
  title={{{MLIR}}: Scaling Compiler Infrastructure for Domain Specific Computation},
  year={2021},
  volume={},
  number={},
  pages={2-14},
  doi={10.1109/CGO51591.2021.9370308}
}

@INPROCEEDINGS{laa_performance_overhead,
  author={Bolat, Alperen and Siddiqui, Fahad and Sezer, Sakir and Tasdemir, Kasim and Khan, Rafiullah},
  booktitle={2023 IEEE 36th International System-on-Chip Conference (SOCC)},
  title={Investigation of Communication Overhead of SoC Lookaside Accelerators},
  year={2023},
  volume={},
  number={},
  pages={1-6},
  doi={10.1109/SOCC58585.2023.10257113}}

@misc{intel_tpp_mlir,
    title={Towards a high-performance AI compiler with upstream MLIR},
    author={Renato Golin and Lorenzo Chelini and Adam Siemieniuk and Kavitha Madhu and Niranjan Hasabnis and Hans Pabst and Evangelos Georganas and Alexander Heinecke},
    year={2024},
    eprint={2404.15204},
    archivePrefix={arXiv},
    primaryClass={cs.PL},
    url={https://arxiv.org/abs/2404.15204}
}

@article{redmule,
title = {RedMule: A mixed-precision matrix–matrix operation engine for flexible and energy-efficient on-chip linear algebra and TinyML training acceleration},
journal = {Future Generation Computer Systems},
volume = {149},
pages = {122-135},
year = {2023},
issn = {0167-739X},
doi = {https://doi.org/10.1016/j.future.2023.07.002},
url = {https://www.sciencedirect.com/science/article/pii/S0167739X23002546},
author = {Yvan Tortorella and Luca Bertaccini and Luca Benini and Davide Rossi and Francesco Conti}
}

@article{nvidiah100,
  title={Nvidia hopper h100 gpu: Scaling performance},
  author={Choquette, Jack},
  journal={IEEE Micro},
  volume={43},
  number={3},
  pages={9--17},
  year={2023},
  publisher={IEEE}
}

@INPROCEEDINGS{gemmini, author={Genc, Hasan and Kim, Seah and Amid, Alon and Haj-Ali, Ameer and Iyer, Vighnesh and Prakash, Pranav and Zhao, Jerry and Grubb, Daniel and Liew, Harrison and Mao, Howard and Ou, Albert and Schmidt, Colin and Steffl, Samuel and Wright, John and Stoica, Ion and Ragan-Kelley, Jonathan and Asanovic, Krste and Nikolic, Borivoje and Shao, Yakun Sophia}, booktitle={2021 58th ACM/IEEE Design Automation Conference (DAC)}, title={Gemmini: Enabling Systematic Deep-Learning Architecture Evaluation via Full-Stack Integration}, year={2021}, volume={}, number={}, pages={769-774}, doi={10.1109/DAC18074.2021.9586216}}

@article{rooflinemodel,
author = {Williams, Samuel and Waterman, Andrew and Patterson, David},
title = {Roofline: an insightful visual performance model for multicore architectures},
year = {2009},
issue_date = {April 2009},
publisher = {Association for Computing Machinery},
address = {New York, NY, USA},
volume = {52},
number = {4},
issn = {0001-0782},
url = {https://doi.org/10.1145/1498765.1498785},
doi = {10.1145/1498765.1498785},
journal = {Commun. ACM},
month = {apr},
pages = {65–76},
numpages = {12}
}

@techreport{amx,
  author      = {Intel Corporation},
  title       = "Accelerate Artificial Intelligence (AI) Workloads with Intel Advanced Matrix Extensions (Intel AMX)",
  institution = "Intel Corporation",
  year        = "2024",
  note        = "version 2024-06-17"
}

@article{rocket,
  title={The rocket chip generator},
  author={Asanovic, Krste and Avizienis, Rimas and Bachrach, Jonathan and Beamer, Scott and Biancolin, David and Celio, Christopher and Cook, Henry and Dabbelt, Daniel and Hauser, John and Izraelevitz, Adam and others},
  journal={EECS Department, University of California, Berkeley, Tech. Rep. UCB/EECS-2016-17},
  volume={4},
  pages={6--2},
  year={2016}
}

@misc{nvdla,
  title = {NVIDIA Deep Learning Accelerator},
  howpublished = {\url{https://nvdla.org/}},
  note = {Accessed: 2024-08-2}
}

@misc{spike,
  title = {Spike, a RISC-V ISA Simulator},
  howpublished = {\url{https://github.com/riscv-software-src/riscv-isa-sim}},
  note = {Accessed: 2025-3-12}
}

@INPROCEEDINGS{crypto_interface,
  author={Hodjat, A. and Verbauwhede, I.},
  booktitle={Conference Record of the Thirty-Eighth Asilomar Conference on Signals, Systems and Computers, 2004.},
  title={Interfacing a high speed crypto accelerator to an embedded CPU},
  year={2004},
  volume={1},
  number={},
  pages={488-492 Vol.1},
  doi={10.1109/ACSSC.2004.1399180}}

@INPROCEEDINGS{crypto_fhe,
  author={Nabeel, Mohammed and Soni, Deepraj and Ashraf, Mohammed and Gebremichael, Mizan Abraha and Gamil, Homer and Chielle, Eduardo and Karri, Ramesh and Sanduleanu, Mihai and Maniatakos, Michail},
  booktitle={2023 Design, Automation \& Test in Europe Conference \& Exhibition (DATE)},
  title={CoFHEE: A Co-processor for Fully Homomorphic Encryption Execution},
  year={2023},
  volume={},
  number={},
  pages={1-2},
  doi={10.23919/DATE56975.2023.10137020}}

@Article{crypto_facv,
AUTHOR = {Gomes, Tiago and Sousa, Pedro and Silva, Miguel and Ekpanyapong, Mongkol and Pinto, Sandro},
TITLE = {FAC-V: An FPGA-Based AES Coprocessor for RISC-V},
JOURNAL = {Journal of Low Power Electronics and Applications},
VOLUME = {12},
YEAR = {2022},
NUMBER = {4},
ARTICLE-NUMBER = {50},
URL = {https://www.mdpi.com/2079-9268/12/4/50},
ISSN = {2079-9268},
DOI = {10.3390/jlpea12040050}
}

@inproceedings{rpc_optimus_prime,
author = {Pourhabibi, Arash and Gupta, Siddharth and Kassir, Hussein and Sutherland, Mark and Tian, Zilu and Drumond, Mario Paulo and Falsafi, Babak and Koch, Christoph},
title = {Optimus Prime: Accelerating Data Transformation in Servers},
year = {2020},
isbn = {9781450371025},
publisher = {Association for Computing Machinery},
address = {New York, NY, USA},
url = {https://doi.org/10.1145/3373376.3378501},
doi = {10.1145/3373376.3378501},
booktitle = {Proceedings of the Twenty-Fifth International Conference on Architectural Support for Programming Languages and Operating Systems},
pages = {1203–1216},
numpages = {14},
location = {Lausanne, Switzerland},
series = {ASPLOS '20}
}

@inproceedings{rpc_zerializer,
author = {Wolnikowski, Adam and Ibanez, Stephen and Stone, Jonathan and Kim, Changhoon and Manohar, Rajit and Soul\'{e}, Robert},
title = {Zerializer: towards zero-copy serialization},
year = {2021},
isbn = {9781450384384},
publisher = {Association for Computing Machinery},
address = {New York, NY, USA},
url = {https://doi.org/10.1145/3458336.3465283},
doi = {10.1145/3458336.3465283},
booktitle = {Proceedings of the Workshop on Hot Topics in Operating Systems},
pages = {206–212},
numpages = {7},
location = {Ann Arbor, Michigan},
series = {HotOS '21}
}

@INPROCEEDINGS{rpc_morpheus,
  author={Tseng, Hung-Wei and Zhao, Qianchen and Zhou, Yuxiao and Gahagan, Mark and Swanson, Steven},
  booktitle={2016 ACM/IEEE 43rd Annual International Symposium on Computer Architecture (ISCA)},
  title={Morpheus: Creating Application Objects Efficiently for Heterogeneous Computing},
  year={2016},
  volume={},
  number={},
  pages={53-65},
  doi={10.1109/ISCA.2016.15}}

@inproceedings{rpc_protocolbuffer_accelerator,
author = {Karandikar, Sagar and Leary, Chris and Kennelly, Chris and Zhao, Jerry and Parimi, Dinesh and Nikolic, Borivoje and Asanovic, Krste and Ranganathan, Parthasarathy},
title = {A Hardware Accelerator for Protocol Buffers},
year = {2021},
isbn = {9781450385572},
publisher = {Association for Computing Machinery},
address = {New York, NY, USA},
url = {https://doi.org/10.1145/3466752.3480051},
doi = {10.1145/3466752.3480051},
booktitle = {MICRO-54: 54th Annual IEEE/ACM International Symposium on Microarchitecture},
pages = {462–478},
numpages = {17},
location = {Virtual Event, Greece},
series = {MICRO '21}
}

@INPROCEEDINGS{dl_riscv_multiprecision,
  author={He, Zicheng and Sheri, Ao and Li, Qiufeng and Cheng, Quan and Yu, Hao},
  booktitle={2023 28th Asia and South Pacific Design Automation Conference (ASP-DAC)},
  title={Agile Hardware and Software Co-design for RISC-V-based Multi-precision Deep Learning Microprocessor},
  year={2023},
  volume={},
  number={},
  pages={490-495},
  doi={}}

@inproceedings{dl_elastic_simd,
author = {Zhang, Zhongcheng and Ou, Yan and Liu, Ying and Wang, Chenxi and Zhou, Yongbin and Wang, Xiaoyu and Zhang, Yuyang and Ouyang, Yucheng and Shan, Jiahao and Wang, Ying and Xue, Jingling and Cui, Huimin and Feng, Xiaobing},
title = {Occamy: Elastically Sharing a SIMD Co-processor across Multiple CPU Cores},
year = {2023},
isbn = {9781450399180},
publisher = {Association for Computing Machinery},
address = {New York, NY, USA},
url = {https://doi.org/10.1145/3582016.3582046},
doi = {10.1145/3582016.3582046},
booktitle = {Proceedings of the 28th ACM International Conference on Architectural Support for Programming Languages and Operating Systems, Volume 3},
pages = {483–497},
numpages = {15},
location = {Vancouver, BC, Canada},
series = {ASPLOS 2023}
}

@INPROCEEDINGS{dl_sssr,
  author={Scheffler, Paul and Zaruba, Florian and Schuiki, Fabian and Hoefler, Torsten and Benini, Luca},
  booktitle={2021 Design, Automation \& Test in Europe Conference \& Exhibition (DATE)},
  title={Indirection Stream Semantic Register Architecture for Efficient Sparse-Dense Linear Algebra},
  year={2021},
  volume={},
  number={},
  pages={1787-1792},
  doi={10.23919/DATE51398.2021.9474230}}

@article{dl_hwacha,
  title={The hwacha microarchitecture manual, version 3.8},
  author={Lee, Yunsup and Ou, Albert and Schmidt, Colin and Karandikar, Sagar and Mao, Howard and Asanovic, K},
  journal={EECS Department, University of California, Berkeley, Tech. Rep. UCB/EECS-2015-263},
  year={2015}
}

@inproceedings{dl_piperench,
author = {Goldstein, Seth Copen and Schmit, Herman and Moe, Matthew and Budiu, Mihai and Cadambi, Srihari and Taylor, R. Reed and Laufer, Ronald},
title = {PipeRench: a co/processor for streaming multimedia acceleration},
year = {1999},
isbn = {0769501702},
publisher = {IEEE Computer Society},
address = {USA},
url = {https://doi.org/10.1145/300979.300982},
doi = {10.1145/300979.300982},
booktitle = {Proceedings of the 26th Annual International Symposium on Computer Architecture},
pages = {28–39},
numpages = {12},
location = {Atlanta, Georgia, USA},
series = {ISCA '99}
}

@INPROCEEDINGS{dl_marsellus,
  author={Conti, Francesco and Rossi, Davide and Paulin, Gianna and Garofalo, Angelo and Di Mauro, Alfio and Rutishauer, Georg and Ottavi, Gian marco and Eggimann, Manuel and Okuhara, Hayate and Huard, Vincent and Montfort, Olivier and Jure, Lionel and Exibard, Nils and Gouedo, Pascal and Louvat, Mathieu and Botte, Emmanuel and Benini, Luca},
  booktitle={2023 IEEE International Solid-State Circuits Conference (ISSCC)},
  title={22.1 A 12.4TOPS/W @ 136GOPS AI-IoT System-on-Chip with 16 RISC-V, 2-to-8b Precision-Scalable DNN Acceleration and 30\%-Boost Adaptive Body Biasing},
  year={2023},
  volume={},
  number={},
  pages={21-23},
  doi={10.1109/ISSCC42615.2023.10067643}}

@INPROCEEDINGS{dl_siracusa,
  author={Scherer, Moritz and Eggimann, Manuel and Mauro, Alfio Di and Prasad, Arpan Suravi and Conti, Francesco and Rossi, Davide and Gómez, Jorge Tomás and Li, Ziyun and Sarwar, Syed Shakib and Wang, Zhao and Salvo, Barbara De and Benini, Luca},
  booktitle={ESSCIRC 2023- IEEE 49th European Solid State Circuits Conference (ESSCIRC)},
  title={Siracusa: A Low-Power On-Sensor RISC-V SoC for Extended Reality Visual Processing in 16nm CMOS},
  year={2023},
  volume={},
  number={},
  pages={217-220},
  doi={10.1109/ESSCIRC59616.2023.10268718}}

@article{dl_riscv_esp,
  title={Enabling heterogeneous, multicore soc research with RISC-V and ESP},
  author={Zuckerman, Joseph and Mantovani, Paolo and Giri, Davide and Carloni, Luca P},
  journal={arXiv preprint arXiv:2206.01901},
  year={2022}
}

@ARTICLE{neuromorphic_neurorvcore,
  author={Yang, Zhijie and Wang, Lei and Shi, Wei and Wang, Yao and Tie, Junbo and Wang, Feng and Yu, Xiang and Peng, Linghui and Xiao, Chao and Xiao, Xun and Yao, Yao and Zhou, Gan and Yu, Xuhu and Gong, Rui and Zhao, Xia and Tang, Yuhua and Xu, Weixia},
  journal={IEEE Transactions on Parallel and Distributed Systems},
  title={Back to Homogeneous Computing: A Tightly-Coupled Neuromorphic Processor With Neuromorphic ISA},
  year={2023},
  volume={34},
  number={11},
  pages={2910-2927},
  doi={10.1109/TPDS.2023.3307408}}

@inproceedings{robotics_racod,
author = {Bakhshalipour, Mohammad and Ehsani, Seyed Borna and Qadri, Mohamad and Guri, Dominic and Likhachev, Maxim and Gibbons, Phillip B.},
title = {RACOD: algorithm/hardware co-design for mobile robot path planning},
year = {2022},
isbn = {9781450386104},
publisher = {Association for Computing Machinery},
address = {New York, NY, USA},
url = {https://doi.org/10.1145/3470496.3527383},
doi = {10.1145/3470496.3527383},
booktitle = {Proceedings of the 49th Annual International Symposium on Computer Architecture},
pages = {597–609},
numpages = {13},
location = {New York, New York},
series = {ISCA '22}
}

@article{aurora,
  author={Kim, Seah and Zhao, Jerry and Asanović, Krste and Nikolić, Borivoje and Shao, Yakun Sophia},
  journal={IEEE Micro},
  title={AuRORA: A Full-Stack Solution for Scalable and Virtualized Accelerator Integration},
  year={2024},
  volume={44},
  number={4},
  pages={97-105},
  doi={10.1109/MM.2024.3409546}
}

@inproceedings{memssa,
  title={Memory SSA-a unified approach for sparsely representing memory operations},
  author={Novillo, Diego and others},
  booktitle={Proceedings of the GCC Developers’ Summit},
  pages={97--110},
  year={2007},
  organization={Citeseer}
}

@article{baccarani2005generalized,
  title={Generalized scaling theory and its application to a $1/4$ micrometer MOSFET design},
  author={Baccarani, Giorgio and Wordeman, Matthew R and Dennard, Robert H},
  journal={IEEE Transactions on Electron Devices},
  volume={31},
  number={4},
  pages={452--462},
  year={2005},
  publisher={IEEE}
}

@article{dennard2007design,
  title={Design of ion-implanted MOSFET's with very small physical dimensions},
  author={Dennard, Robert H and Gaensslen, Fritz H and Yu, Hwa-Nien and Rideovt, V Leo and Bassous, Ernest and Leblanc, Andre R},
  journal={IEEE Solid-State Circuits Society Newsletter},
  volume={12},
  number={1},
  pages={38--50},
  year={2007},
  publisher={IEEE}
}

@article{danowitz2012cpudb,
author = {Danowitz, Andrew and Kelley, Kyle and Mao, James and Stevenson, John P. and Horowitz, Mark},
title = {CPU DB: recording microprocessor history},
year = {2012},
issue_date = {April 2012},
publisher = {Association for Computing Machinery},
address = {New York, NY, USA},
volume = {55},
number = {4},
issn = {0001-0782},
url = {https://doi.org/10.1145/2133806.2133822},
doi = {10.1145/2133806.2133822},
journal = {Commun. ACM},
month = apr,
pages = {55–63},
numpages = {9}
}

@INPROCEEDINGS{horowitz2014computing,
  author={Horowitz, Mark},
  booktitle={2014 IEEE International Solid-State Circuits Conference Digest of Technical Papers (ISSCC)},
  title={1.1 Computing's energy problem (and what we can do about it)},
  year={2014},
  volume={},
  number={},
  pages={10-14},
  doi={10.1109/ISSCC.2014.6757323}
}

@article{logca,
author = {Altaf, Muhammad Shoaib Bin and Wood, David A.},
title = {LogCA: A High-Level Performance Model for Hardware Accelerators},
year = {2017},
issue_date = {May 2017},
publisher = {Association for Computing Machinery},
address = {New York, NY, USA},
volume = {45},
number = {2},
issn = {0163-5964},
url = {https://doi.org/10.1145/3140659.3080216},
doi = {10.1145/3140659.3080216},
journal = {SIGARCH Comput. Archit. News},
month = jun,
pages = {375–388},
numpages = {14}
}
\end{multicols}
\twocolumn

\clearpage
\appendix

\section{Artifact Appendix}

\subsection{Abstract}

This artifact serves to validate the implementation and evaluation sections of the above work.
The artifact is in the form of a GitHub Repository plus a Docker Container for bundled pre-build binary dependencies.
No special hardware or powerful computers are needed for evaluation, and a full reproduction of the results should take less than two hours.
The artifact consists of tiled matrix multiplication code written in C and MLIR, that is compiled with various levels of optimization to two RISC-V based architectures (Gemmini~\cite{gemmini} and OpenGemm~\cite{opengemm}). Simulators for these architectures are included in the Docker image.
Notably, the optimizations presented in \autoref{sec:implementation} are performed and evaluated.
This artifact servres to reproduce the main claims of the paper and all figures of the evaluation section.

\subsection{Artifact check-list (meta-information)}

{\small
\begin{itemize}
  \item {\bf Algorithm: } No novel algorithm.
  \item {\bf Program: } Tiled Matrix Multiplication.
  \item {\bf Compilation: } Custom RV32 and RV64 compilers included.
  \item {\bf Transformations: } Yes, it's all included.
  \item {\bf Binary: } x86 binaries are included.
  \item {\bf Model: } No model used.
  \item {\bf Data set: } No data set used.
  \item {\bf Run-time environment: } Linux/Docker.
  \item {\bf Hardware: } No specific hardware needed.
  \item {\bf Run-time state: } Not sensitive.
  \item {\bf Execution: } Less than 2hrs, no special care needed.
  \item {\bf Metrics: } Instruction/Cycle counts, bandwidth, throughput.
  \item {\bf Output: } Console prints, Figures, Data tables.
  \item {\bf Experiments: } Docker and scripts.
  \item {\bf How much disk space required (approximately)?: } 16 GB.
  \item {\bf How much time is needed to prepare workflow (approximately)?: } 5 Minutes.
  \item {\bf How much time is needed to complete experiments (approximately)?: } 2 Hours.
  \item {\bf Publicly available?: } Yes.
  \item {\bf Code licenses: } Apache v2.0 with LLVM Exceptions.
  \item {\bf Archived: } \href{https://doi.org/10.5281/zenodo.16260752}{10.5281/zenodo.16260752}
\end{itemize}
}

\subsection{Description}

\subsubsection{How to access}

Get it from Zenodo\footnote{\url{https://doi.org/10.5281/zenodo.16260752}} or clone from GitHub \url{https://github.com/KULeuven-MICAS/accfg-artifacts}.

\subsubsection{Hardware dependencies}
These experiments should run on any typical x86 computer. As all hardware testing is done in simulation, no special hardware is required. Some parts of those simulations/postprocessing are rather compute-intensive, and will run faster on a faster machine. All software needs to be pulled from ghcr.io and pypi, machines with faster internet connection will finish this faster. These requirements should be plenty:
\begin{itemize}
\item    4-core x86 CPU
\item    8 GB RAM
\item    A working internet connection
\item    16 GB of free disk space
\end{itemize}

\subsubsection{Software dependencies}

\begin{itemize}
  \item Linux (to run docker) or other linux-like environments
  \item Git (to obtain the source code for the experiments)
  \item Docker (to run the experiments)
\end{itemize}

\subsection{Installation}

Get the repository by running \texttt{git clone --recursive https://github.com/KULeuven-MICAS/accfg-artifacts} or by extracting the repo tarball from Zenodo. When getting the files from Zenodo, the image can be found alongside the repository and can be imported using \texttt{docker image load accfg-artifacts.container.tar.gz}.

\subsubsection{Basic Test}

A basic test can be performed by cloning the repo (with submodules) and then running the following command: \texttt{docker run --rm -itv \$PWD:/repo:z ghcr.io/} \texttt{kuleuven-micas/accfg-artifacts:latest}\linebreak \texttt{/repo/hello-world.sh} (with no space between ghcr.io/ and kuleuven). This command is also documented in the README, where it can be more easily copied from.

\subsection{Experiment workflow}

See the README for more detailed information on the flow and how to interpret the results. For a full run, use \texttt{./run-all.sh}. (This requires rootless docker. If docker is not rootless, use \texttt{sudo ./run-all.sh}).

\textbf{Note on time:} On modern Desktop CPUs we saw times at around ~15 minutes to complete the entire evaluation, but lower-power CPUs seem to struggle a lot with the simulation workloads. We based our time estimations on typical student machine. If there is access to a fast workstation, times can be expected to be significantly lower.

\subsection{Evaluation and expected results}

\textbf{Claims:}

\begin{itemize}
  \item Performance on OpenGEMM is improved by 1.99x geomean, and up to 2.71x for some sizes through our optimizations.
  \item Performance on Gemmini is improved by 10.5\% geomean.
  \item \autoref{fig:gemmini_performance}, \autoref{fig:snax_measured_perf} and \autoref{fig:snax_roofline_plot} in the paper can be reproduced.
\end{itemize}

See the included README for additional guidance on how to interpret the results.

\subsection{Notes}

The projects README file has a lot of additional information and notes, and provides additional guidance for performing individual steps.

%
%

\end{document}